\documentclass[preprint,floatfix,groupedaddress,nofootinbib]{revtex4-2}

\usepackage{algorithm}
\usepackage{algpseudocode}
\usepackage{amsmath}
\usepackage{amsthm}
\usepackage{amssymb}
\usepackage{amscd}
\usepackage{bm}
\usepackage{booktabs}
\usepackage{latexsym}
\usepackage{mathtools}
\usepackage{comment}

\usepackage{graphicx}
\usepackage{epsfig}
\usepackage{epstopdf}
\usepackage{svg}
\usepackage[caption=false]{subfig}
\usepackage{placeins}
\usepackage{tabularx}

\usepackage{url}
\usepackage{hyperref}

\providecommand{\about}[0]{\raise.17ex\hbox{$\scriptstyle\sim$}}

\providecommand{\eqref}[1]{(\ref{#1})}

\providecommand{\vanish}[1]{}

\usepackage{multirow} 

\graphicspath{ {./figures/} }

\begin{document}
	
\title{Dependence of simulated radiation damage on crystal structure and atomic misfit in metals}

\author{J. C. Stimac}
\affiliation{Department of Chemical Engineering, University of California, Davis, CA, 95616, USA.}
\affiliation{Lawrence Livermore National Laboratory, Livermore, CA, 94550, USA.}

\author{C. Serrao}
\affiliation{Department of Materials Science and Engineering, University of California, Davis, CA, 95616, USA.}

\author{J. K. Mason}
\email{jkmason@ucdavis.edu}
\affiliation{Department of Materials Science and Engineering, University of California, Davis, CA, 95616, USA.}

\begin{abstract}
This study investigates radiation damage in three metals in the low temperature and high radiant flux regime using molecular dynamics and a Frenkel pair accumulation method to simulate up to $2.0$ displacements per atom.
The metals considered include Fe, equiatomic CrCoNi, and a fictitious metal with identical bulk properties to the CrCoNi composed of a single atom type referred to as an A-atom.
CrCoNi is found to sustain higher concentrations of dislocations than either the Fe or A-atom systems and more stacking faults than the A-atom system.
The results suggest that the concentration of vacancies and interstitials are substantially higher for the CrCoNi than the A-atom system, perhaps reflecting that the recombination radius is smaller in CrCoNi due to the roughened potential energy landscape.
A model that partitions the major contributions from defects to the stored energy is described, and serves to highlight a general need for higher fidelity approaches to point defect identification. 
\end{abstract}


\maketitle

\section{Introduction}
\label{sec:introduction}

Structural components in nuclear fission reactors need to be engineered to withstand decades of exposure to radiation and elevated temperatures \cite{zinkle2009structural,was2016fundamentals}.
Materials in the next generation of fission reactors and future fusion reactors will be subject to substantially higher radiation dosages and temperatures and likely highly corrosive environments as well \cite{geniv2002,geniv2020}.
Irradiation by exposure to high-energy particles displaces atoms and damages the crystalline structure of pure metals and alloys \cite{nordlund2018improving,nordlund2018primary}, resulting in a variety of microstructural changes.
Specifically with respect to the degrading effects of radiation, there are five considerations for structural metals in reactor environments:
radiation hardening (low temperature), radiation-induced segregation and precipitation, void swelling, radiation-induced creep, and helium embrittlement (high temperature) \cite{2006schilling,2013zinkle}.
The three intermediate temperature effects are usually the most relevant in practice, and are often observed simultaneously since they are all strongly associated with the underlying ability of point defects generated by radiation to migrate through the lattice \cite{mansur1994theory,ardell2016radiation}.

Radiation-induced segregation of substitutional solutes is generally attributed to the inverse Kirkendall effect where, as vacancies migrate to and annihilate on sinks, the different atomic species have different migration rates in the opposite direction \cite{anthony1969solute}.
The segregation extent is governed by differences in diffusive mobility, with undersized solutes generally being enriched and oversized solutes being depleted in the vicinity of the sinks \cite{ardell2008radiation,ardell2016radiation}.
Void swelling is a serious engineering concern that entails an increase in the volume of the material by as much as $1\%$ per dpa \cite{garner2000comparison} from the nucleation and growth of voids in the bulk.
This process is driven by an imbalance in the concentrations of vacancies and interstitials, with annihilation of high-mobility interstitials on dislocations and other sinks leaving behind a relative excess of vacancies that precipitate as voids \cite{mansur1994theory,taller2021predicting}.
While a variety of mechanisms have been proposed for irradiation creep, the two main mechanisms are believed to be the climb of favorably-oriented dislocations by the stress-induced preferred absorption of point defects and the glide of dislocations enabled by climb over obstacles \cite{matthews1988irradiation,mansur1994theory}.


Multi-principal component alloys (MPEAs) \cite{2004cantor,2004yeh} (often called high-entropy alloys or compositionally complex alloys) consist of a few to several atom types in solid solution with nominally equi-atomic concentrations.
This emerging class of metals could perform well as structural materials in irradiated environments, with initial evidence showing exceptional mechanical properties \cite{gludovatz2014fracture,george2019high} and higher resistance to radiation damage \cite{granberg2016mechanism,zhang2022effects} than traditional metals.
One experimental and computational study of the effects of radiation on a CrCoNi MPEA showed decreases in the relative disorder, the number of defects, and large defect clusters compared to irradiated samples of pure Ni and a NiFe binary alloy \cite{granberg2016mechanism}.
The authors attributed these effects to decreasing dislocation mobility with increasing number of atomic species.
Lu et al.\ \cite{lu2019irradiation} studied void swelling in several Ni-based alloys including pure Ni, NiFe, CrCoNi and two quinary alloys
with and without prior nanoindentation, and found that nanoindentation improved resistance to swelling by increasing the density of defects like dislocations, stacking faults, and twin boundaries that promote vacancy annihilation.
Curiously, Veli{\c{s}}a et al.\ \cite{velicsa2020temperature} found that CrCoNi showed superior irradiation resistance relative to NiCr only at temperatures near and below $300 \ \mathrm{K}$ though.
The reason for this is not well established, but is believed to be related to the specifics of the chemical short range order (SRO) that developed in the two alloys.

The most direct way to simulate radiation damage at the atomic level uses molecular dynamics (MD) simulations of collision cascades \cite{nordlund2018primary,nordlund2018improving}.
This involves assigning a large initial velocity to the primary knock-on atom (PKA) to mimic a passing neutron or other high-energy particle transferring kinetic energy to the lattice.
The PKA recoils, displacing many of the surrounding atoms from their lattice sites and converting the initial kinetic energy into a thermal spike with sufficient energy to facilitate the regeneration of the crystalline lattice and the recombination of many, but not all, of the generated point defects.
The resulting interstitials and vacancies are respectively distributed on the periphery and the interior of the affected zone, increase the point defect concentration in the material, and are directly responsible for the most visible degrading effects of radiation.

Mass conservation requires that the interstitials and vacancies generated by a single collision cascade occur as Frenkel pairs.
The number of Frenkel pairs generated in this way divided by the number of atoms in the material is known as the displacements per atom (dpa), and is the standardized measure of the extent of radiation damage in crystalline materials \cite{astm2016standard}.
Early work by Kinchin and Pease modeled atoms as hard-spheres that exhibit elastic collisions during collision cascades and laid the theoretical foundation for radiation effects in crystalline materials \cite{kinchin1955displacement}.
If the energy transferred to an atom exceeded a material-specific threshold value, then the atom was said to have been displaced from its lattice site.
While energies below the threshold could still displace the atom, it would return to its lattice site after the initial perturbation.
The theory developed by Kinchin and Pease was further developed in the work of Norgett, Robinson, and Torrens (hereafter referred to as the NRT model) who added additional terms to account for energy lost to ionization and for the affects of inelastic collisions \cite{norgett1975proposed}.
Several more recent studies concluded that the NRT model overestimates the number of defects generated by a collision cascade and neglects the mixing from atomic replacements though, spurring a number of proposed refinements \cite{nordlund2018primary,nordlund2018improving}.
These are significant for the reason that accurate damage models that can reliably predict dpa are essential to reliably compare radiation damage resistance among various materials.

Full-scale atomic simulations of collision cascades have been performed for several decades now, and are useful to uncover the evolution of radiation-induced primary damage \cite{nordlund1995molecular,beardmore1998efficient}.
However, the use of MD for this application is subject to several limitations.
As Ref.\ \cite{nordlund2018primary} points out, interatomic potentials cannot capture the effects of deviations from the Born-Oppenheimer approximation when excited electronic states are induced.
The other main limitations are the time and length scales that can reasonably be achieved with modern computational resources.
MD simulations are usually no longer than a few nanoseconds, limiting the overall radiation dose that can reasonably be achieved by full cascade simulations without the events overlapping in time;
Refs.\ \cite{Derlet2020,Chen2020} further discuss these limitations.
As a result of the high computational cost to reach appreciable dpa, the doses investigated in MD simulations have been historically been less than $1.0$ dpa.
Although these are useful to understand defect creation at low doses, structural materials in a nuclear reactor core can experience as much as $80$ dpa over a $40$ year service life \cite{2013zinkle}.
 
A variety of strategies have been used to circumvent the computational limitations imposed by cascade simulations, one of which involves the direct insertion of a high density of Frenkel pairs \cite{crocombette2006atomistic,belko2006frenkel,chartier2008frenkel}.
Such Frenkel pair accumulation (FPA) techniques forgo the dynamics of time-resolved high-energy atomic collisions stemming from the primary knock-on event.
Instead, they intermittently introduce Frenkel pairs by randomly displacing atoms from their lattice positions, usually followed by some form of equilibration and time integration.
These methods benefit from a clearly defined radiation dose, based on the the number of displaced atoms, and dose rate, based on the ratio of displaced atoms to the simulated time.
For example, Chartier et al.\ \cite{chartier2016early} used a FPA procedure to model irradiation of $\mathrm{UO_2}$ and found the steady-state dislocation density to be in good agreement with experiments.
A major limitation of FPA procedures though is the absence of any effects related to the thermal spike \cite{Chen2020}, particularly providing the thermal energy necessary for diffusion and clustering of defects;
this includes the recombination of point defects when vacancies and interstitials collide.
Analysis of FPA simulations should therefore be done with careful consideration of this limitation to avoid the potential for nonphysical extrapolation.
 
Recent work by Derlet and Dudarev \cite{Derlet2020} introduced a variant of the FPA method that further streamlines the process of sampling irradiated microstructures.
Known as the creation-relaxation algorithm (CRA), this differs from the preceding FPA methods in that there are no time-integrated dynamics.
CRA simulations randomly select atoms and displace them with random directions and magnitudes, just as other FPA methods do, but always follow this by potential energy minimization.
The entire simulation involves repeating this process for a specified number of displacements, with the canonical dpa equal to the number of displaced atoms divided by the total number of atoms in the system.
Dudarev and Derlet applied the CRA to BCC Fe systems of a variety of sizes and reported good agreement between full cascade simulations \cite{byggmastar2018effects,granberg2019cascade} and the CRA for interstitial density as a function of dpa, at least up to a linear rescaling of both the independent and dependent variables.
The need for rescaling is likely related to the CRA effectively being performed at zero Kelvin  \cite{chartier2019rearrangement};
the thermally-driven diffusion of point defects that is prevalent at high temperatures and substantially contributes to microstructure evolution of irradiated materials is negligible in such conditions.
That said, the CRA can be viewed as simulating radiation damage in conditions where thermally-driven diffusion is active but negligible compared to other mass transport mechanisms.

 
The CRA has since been used to investigate radiation effects in materials other than BCC Fe.
One such study used the CRA to simulate irradiation of a NiFe system doped with carbon and evaluated the ability of carbon interstitials to decrease radiation damage \cite{ge2020effects}.
Several others applied the CRA to tungsten to investigate the relationships between dpa and specific physical parameters at relatively high doses (above $1.0$ dpa), either alone or with the assistance of other computational or experimental methods.
Parameters that were investigated include thermal conductivity \cite{mason2021estimate}, hydrogen embrittlement and tritium concentration \cite{mason2021parameter}, and radiation induced structural evolution \cite{mason2020observation}.

This paper investigates the microstuctures of highly irradiated Fe, equi-atomic CrCoNi, and a fictitious metal with identical bulk properties to the CrCoNi composed of a single atom type referred to as an A-atom.
The main motivation for including the Fe and A-atom systems is to establish points of comparison for the investigation of the reported radiation resistance of CrCoNi.
The CRA is used to simulate the irradiation of all three systems up to a final dose of $2.0$ dpa.
The details of the implemented CRA, as well as an analysis of the experimental dose rates and temperatures for which it is likely relevant, are described in Sec.\ \ref{sec:methods}.
The same section also outlines our methods for identifying material defects including dislocations, vacancies, interstitials, and stacking faults, and a model for the energy stored in those defects. 
The results and a discussion of the simulations are included in Sec.\ \ref{sec:results_and_Discussion}, and Sec.\ \ref{sec:conclusion} draws conclusions to inform further research in this area.

\section{Methods}
\label{sec:methods}

\subsection{CRA simulations}
\label{subsec:cra}

All molecular dynamics simulations were performed using the LAMMPS software \cite{LAMMPS}.
Orthorombic simulation cells were used with periodic boundary conditions for all cell faces, and the number of atoms remained constant.
No temperature or time-steps were defined for any simulation.
After constructing the initial space-filling single crystals, the volume of the simulation cell was relaxed using a potential energy minimization and fixed thereafter.
The present study examined three material systems: BCC Fe, equi-atomic FCC CrCoNi, and FCC A-atom designed to reproduce the bulk properties of CrCoNi using a single fictitious atom type \cite{varvenne2016average}. 
Comparison with an A-atom model more directly allows identification of the effects caused by chemical short range order (SRO) and lattice distortion (LD) which are widely implicated in the enhanced physical properties of MPEAs \cite{george2019high}. 

The BCC Fe simulation consisted of a simulation volume of $40$ units cells along each of the three dimensions, and with two atoms in each unit cell contained a total of $128\ 000$ atoms.
This simulation used the Mendelev-II embedded atom method (EAM) potential \cite{mendelev2003development}.
The FCC CrCoNi simulation consisted of $32$ unit cells along each of the three dimensions, and with four atoms in each unit cell contained a total of $131\ 072$ atoms.
All three atomic species were represented with equal concentrations, and were initially distributed uniformly at random throughout the simulation cell.
The purpose of reducing the number of unit cells for the FCC systems was to make the number of atoms as close as possible to that in the Fe simulation.
The equiatomic CrCoNi simulation utilized the EAM potential developed by Li et al.\ \cite{li2019strengthening}.
The FCC A-atom simulation used the EAM potential developed by Jian et al.\ \cite{jian2020effects} but was otherwise identical to the CrCoNi simulation.

\begin{figure}
	\centering
	\includegraphics[width=0.40\textwidth]{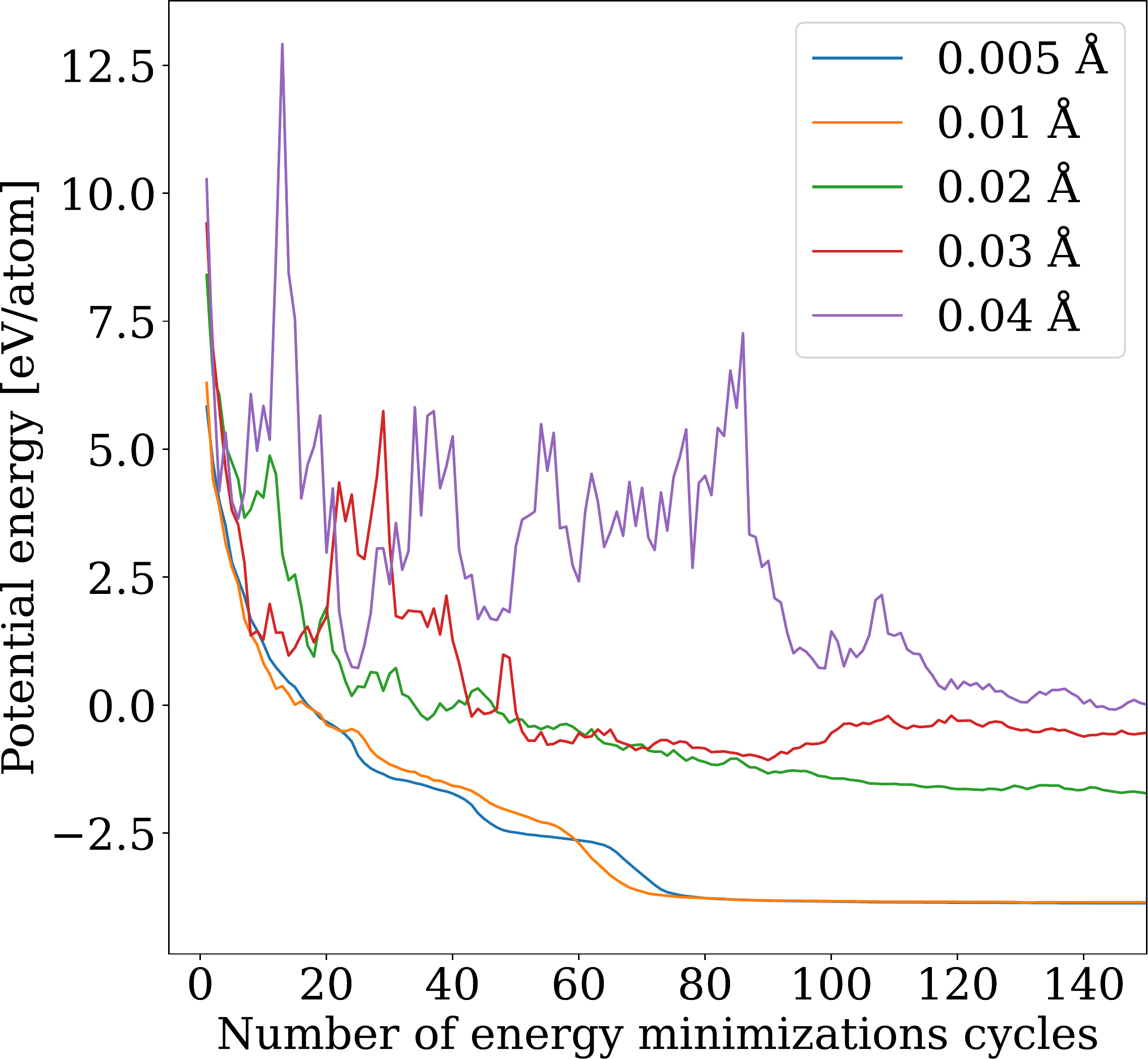}
	\caption{The potential energy per atom of a highly defected BCC Fe system as a function of minimization cycles, measured with respect to the energy of an isolated atom.
	Perturbation magnitudes of $0.005$ and $0.01$ angstroms allowed the system to quickly reach a deep potential energy minimum.}
	\label{fig:shake_test}
\end{figure}

This work circumvented the computational limitations associated with full collision cascade simulations by using the creation relaxation algorithm (CRA) \cite{Derlet2020} to simulate the effects of irradiation by the repeated introduction of Frenkel pairs \cite{Chatier2005,chartier2016early};
the basic CRA is described in Alg.\ \ref{alg:cap}.
All three experiments used the CRA to simulate radiation damage up to $2.0$ dpa, requiring a total of $256\ 000$ atomic displacements in the Fe system and $262\ 144$ displacements in both the CrCoNi and A-atom systems.
The potential energy minimization conducted after each atomic displacement consisted of two steps.
First, the position of every atom was randomly perturbed by a small Gaussian-distributed displacement.
Second, a standard gradient-based energy minimization was applied to the atomic positions to reduce the system energy.
The purpose of the small perturbations in the first step was to help the system escape shallow local energy minima as would naturally occur in a thermal system.
The perturbation magnitude was fixed after systematically exploring the relationship between the perturbation magnitude and the relaxed potential energy of a smaller BCC Fe system of $2000$ atoms.
This involved displacing $100$ atoms and performing energy minimizations without perturbations between displacements.
After all the displacements, the atoms were subjected to repeated cycles of perturbations and energy minimization, with the resulting potential energy profiles for a given standard deviation of the perturbation magnitude shown in Fig.\ \ref{fig:shake_test} as a function of cycle number.
Standard deviations of $0.005$ and $0.01$ angstroms were found to rapidly result in a relatively stable potential energy minimum, with larger displacements often introducing additional defects and smaller displacements unable to reliably allow the system to escape shallow potential energy minima.
A perturbation magnitude of $0.01$ angstroms was chosen for the subsequent simulations for the reason that it reached the potential energy minimum in the fewest number of cycles.

\begin{figure}
\begin{algorithm}[H]
\caption{Creation Relaxation Algorithm}\label{alg:cap}
\begin{algorithmic}[1]
\State $n := 0$ \Comment{number of displaced atoms}
\While{$n / N < \phi_\mathrm{dpa}$} \Comment{$N$ is total number of atoms}
\State Randomly select an atom $i$ uniformly over all atoms 
\State Move atom $i$ to a randomly selected point within the simulation cell
\While{potential energy increase exceeds a threshold}
\State Move atom $i$ to a randomly selected point within the simulation cell
\EndWhile
\State Relax the system using energy minimization 
\State $n \gets n + 1$
\EndWhile 
\end{algorithmic}
\end{algorithm}
\end{figure}

Since the intention was for the perturbations to make the system more closely resemble a thermal one, a natural question is what would be the corresponding temperature for a given perturbation magnitude.
The equipartition theorem implies that the potential energy per atom relative to that in an ideal crystal should be $(3 / 2) k_\mathrm{B} T$ for a thermal system;
equating this with the average potential energy increase per atom resulting from a single perturbation gives an equivalent temperature of $\about 30 \ \mathrm{K}$ for all three systems.
This implies that the additional energy supplied by the perturbations is consistent with the assumption that any thermally-driven diffusion of point defects is insignificant compared to alternative mass transport mechanisms.

In addition to the three simulations described above, another CRA simulation of equiatomic CrCoNi with $64$ unit cells along each dimension and a total of $1\ 048\ 576$ atoms was conducted to $3.0 \ \mathrm{dpa}$ to investigate the effect of system size.
With energy minimization being by far the most computationally intensive step in the CRA, this simulation performed $25$ atomic displacements (step $4$ of Alg.\ \ref{alg:cap}) per energy minimization to reduce the simulation run time;
the larger volume of this simulation decreases the probability that multiple atomic displacements occur in a given region without an intervening relaxation event.
The main results of this simulations are presented in Appendix \ref{sec:appendix}.

\subsection{Applicable temperatures and dose rates}
\label{subsec:temperatures}

As mentioned in Sec.\ \ref{sec:introduction}, the microstructures that develop when using the CRA are a direct product of relaxations of the atomic-level stress fields and do not necessarily represent microstructures that develop in conditions with appreciable thermally-driven diffusion. 
We characterize the regime of physical conditions for which the CRA is applicable by a simple argument that compares the rates of atomic transport by thermally-driven diffusion and ballistic displacements from collision cascades.
The main criterion for applicability is that the rate of ballistic displacements per atom $K_{0}$ be much greater than the interstitial hopping rate $\gamma_{\mathrm{diff}}$ (assumed to be the fastest thermally-driven diffusion event).
In such conditions, any effects of thermally-driven diffusion should be negligible compared to those resulting from ballistic displacements and the subsequent stress relaxation.
The interstitial hopping rate, as described by transition state theory, obeys the Arrhenius relation $\gamma_{\mathrm{diff}} = \nu_0 \mathrm{exp}[-E_\mathrm{a}/(k_B T)]$ where $\nu_0$ is the high-temperature limit for the site-hopping frequency, $E_\mathrm{a}$ is the activation energy for interstitial diffusion, $k_B$ is Boltzmann's constant, and $T$ is the temperature.
The random walk diffusion model suggests that $\nu_0$ be approximated by $6 D_0 / \lambda^2$ where $D_0$ is the diffusivity prefactor and $\lambda$ is the distance separating two interstitial sites \cite{allen2005kinetics}. 

\begin{figure}
	\centering
	\includegraphics[width=0.45\textwidth]{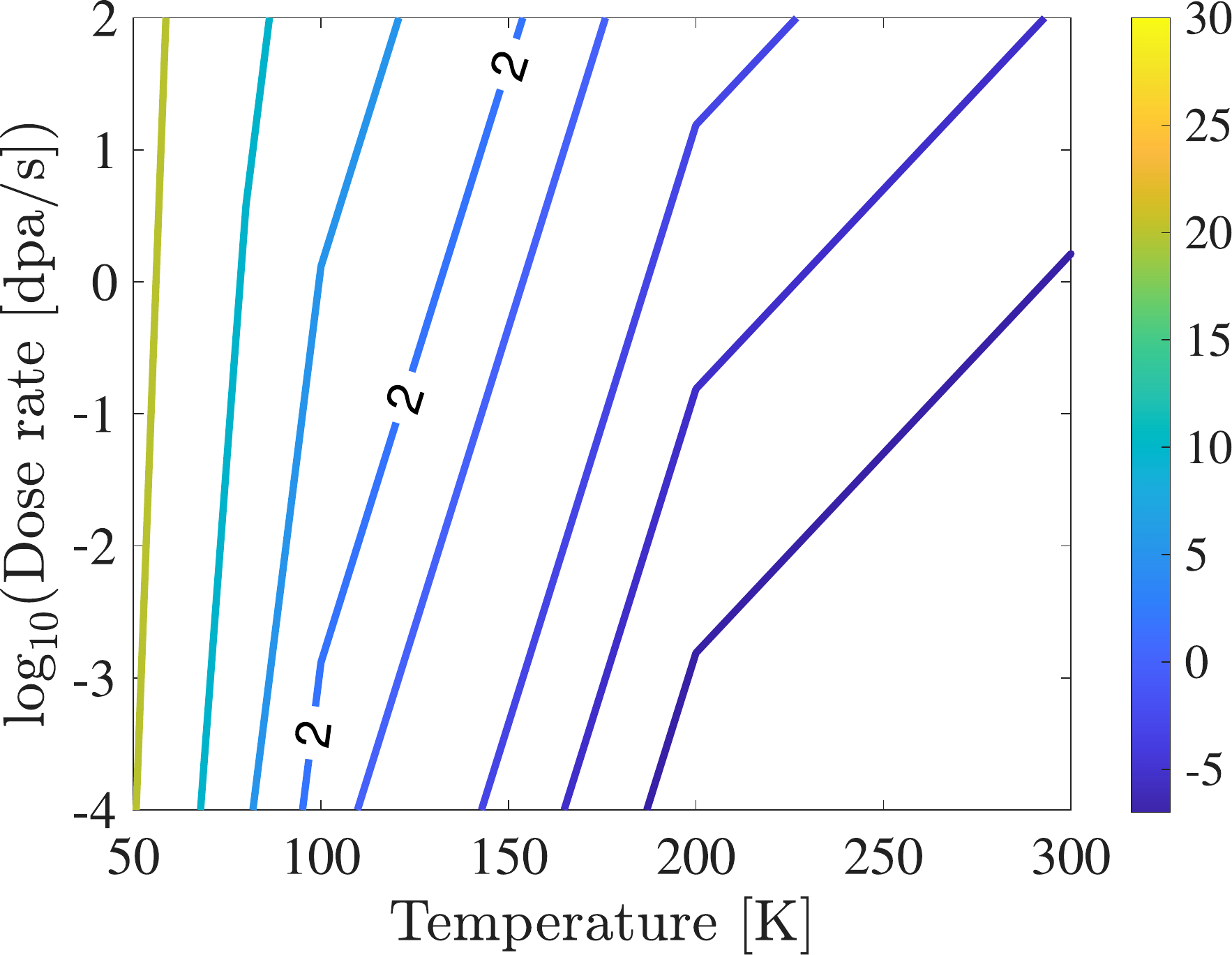}
	\caption{Contour plot of the applicability measure $\mathcal{A} = \log_{10}(K_0 / \gamma_{\mathrm{diff}})$ as a function of the logarithm of the dose rate and temperature.
	The calculations used the diffusivity and activation energy for interstitial diffusion in CrCoNi given by Ref.\ \cite{zhao2017preferential}.}
	\label{fig:applicability_regime}
\end{figure}

Using the values for the prefactor and activation energy for interstitial diffusion for CrCoNi given by Ref.\ \cite{zhao2017preferential} leaves only the temperature and $K_0$ as independent variables.
A measure of applicability $\mathcal{A} = \log_{10} (K_0 / \gamma_{\mathrm{diff}})$ is defined such that an increase of $\mathcal{A}$ by one indicates an order of magnitude increase in the ratio of $K_0$ to $\gamma_{\mathrm{diff}}$.
Figure \ref{fig:applicability_regime} gives a contour plot of $\mathcal{A}$ for CrCoNi and indicates that applicability increases at lower temperatures and higher dose rates;
moreover, the dependence on the temperature is much more significant than the dose rate, a consequence of the exponential scaling in the Arrhenius relation for the diffusivity.
For the dose rates considered, the temperatures for which $\mathcal{A} \ge 2$ are all in the cryogenic range.
This low temperature constraint is entirely consistent with the equivalent temperature resulting from the perturbations in the atomic positions described in Sec.\ \ref{subsec:cra}.

\subsection{Identifying defects}
\label{subsec:identifying}

It is difficult to precisely identify crystalline defects in a material that has suffered extensive radiation damage.
Certainly there is still an underlying lattice, but conventional defect models assume that the defect is isolated, or equivalently, that the surrounding atoms occupy well-defined lattice positions.
This is not true for the high defect concentrations that can occur in irradiated materials, and the identification of reference lattice sites is further complicated in MPEAs like CrCoNi where the reference lattice is already perturbed by atomic size differences.

The numbers and types of dislocations in the simulations were determined using OVITO's dislocation extraction algorithm (DXA) \cite{ovito};
the algorithm requires a trial circuit length and a value for circuit stretchability which were set to $14$ and $9$ respectively.
Surprisingly, the DXA appeared to be relatively robust to extensive disruption of the crystalline lattice from the CRA, with very few isolated dislocation segments appearing in the networks reported below in Sec.\ \ref{subsec:dislocations}.

\begin{figure}
	\centering
	\includegraphics[width=0.45\textwidth]{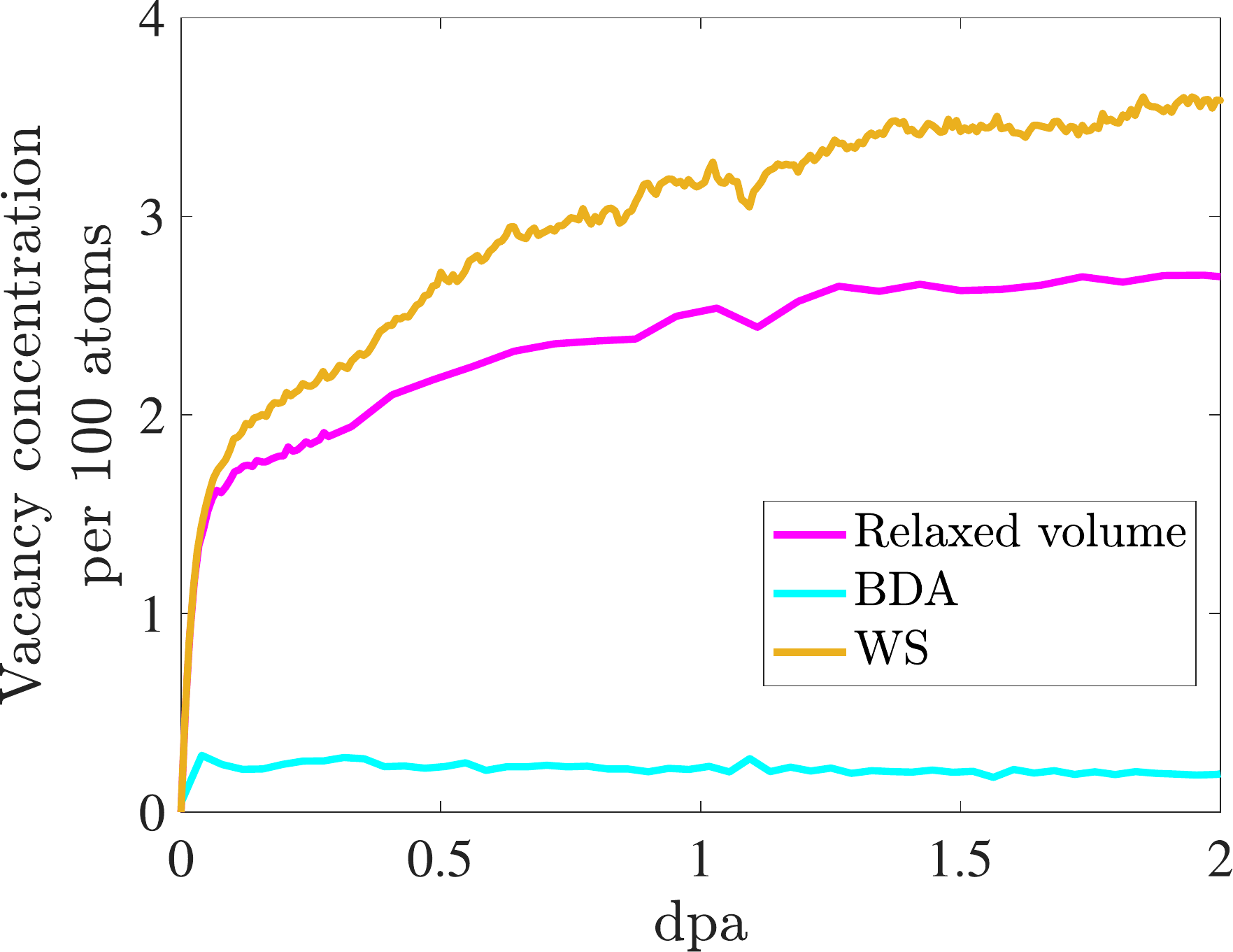}
	\caption{Three different estimates for the number of vacancies in BCC Fe as a function of dpa. The BDA estimate is likely inaccurate due to the difficulty of identifying crystal lattice sites in highly defected materials.}
	\label{fig:est_frankel}
\end{figure}

The number of vacancies were determined using OVITO's Wigner-Seitz (WS) analysis.
This constructs Voronoi cells around the lattice sites in an initial crystal structure and checks the atom occupancy of each Voronoi cell in subsequent time steps.
A Voronoi cell that does not contain any atoms is regarded as indicating the presence of a vacancy, though this approach does not precisely define the vacancy location.
The WS analysis was compared with two other approaches to estimate vacancy concentration in the BCC Fe systems, the first of which was the BCC defect analysis (BDA) \cite{moller2016bda}.
The second approach involved evaluating the relaxed volume of the simulation box.
Inserting a single Frenkel pair produces an interstitial and a lattice site occupied by a vacancy, resulting in the expansion of a crystal subject to zero traction boundary conditions.
Since the simulations were conducted at constant volume, the insertion of Frenkel pairs in the CRA instead elevated the system pressure.
The equivalent volume change was evaluated by relaxing the simulation cell subject to a zero traction boundary condition, and enabled the number of vacancies to be estimated by assuming that the volume change from inserting a Frenkel pair remained constant throughout the simulation.
Figure \ref{fig:est_frankel} compares the three methods and shows that the WS analysis agrees much better with with the vacancy concentration estimated from the relaxed volume than from the BDA. 
The factor by which the WS and relaxed volume method differ can be interpreted as the reduction in volume change per effective Frenkel pair insertion with increasing dpa and is related to point defect interactions.
The poor performance of the BDA is likely a consequence of extensive radiation damage making it difficult to identify vacancies based on features of local atomic environments without a clear underlying crystal lattice.

\begin{figure}
	\centering
    {\includegraphics[width=0.45\textwidth]{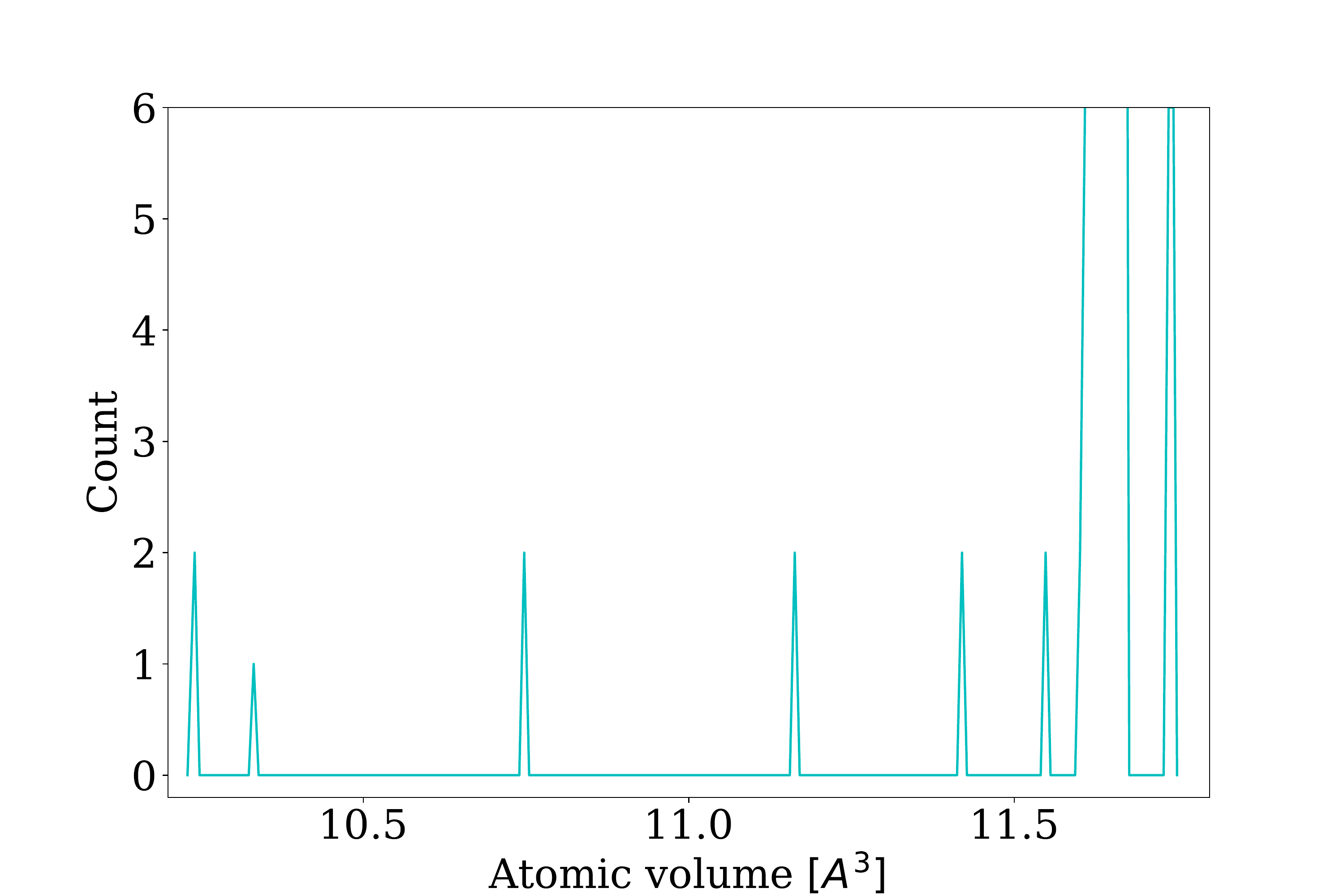}}\ 
	{\includegraphics[width=0.45\textwidth]{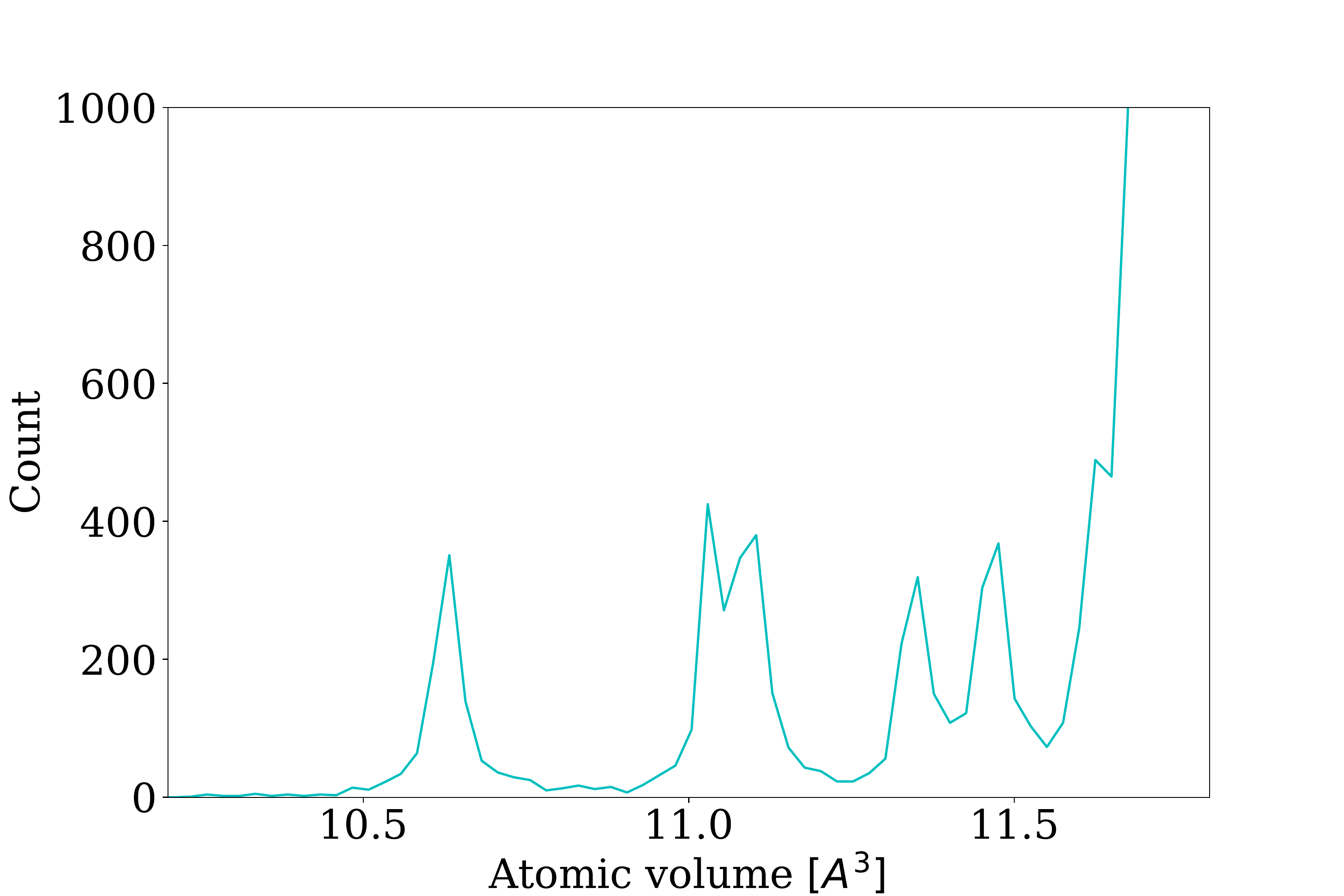}}
	\qquad
	\caption{Atomic volume distribution as given by Voronoi tessellations in BCC Fe for (left) an isolated interstitial and (right) at $0.01$ dpa.
 All histograms were evaluated using $200$ equally-sized bins within the domain of the distribution.
 The simulation cells for the isolated interstitial and the $0.01$ dpa distributions contained $54\ 000$ and $128\ 000$ atoms, respectively.}
	\label{fig:atomic_vol}
\end{figure}

Interstitials are the defect type that is most difficult to identify in our simulations.
The multiplicity of local atomic configurations that can occur for non-isolated interstitials means that template-based approaches would be difficult or even infeasible to implement, and examining the expected number of atoms contained within a surface that passes only through crystalline material (the analogue of a Burgers circuit) often fails because of the obstructions to constructing such surfaces.
In the absence of a canonical alternative, our approach involves constructing a distribution of atomic volumes as estimated by the Voronoi polyhedra.
The compressive stresses around an interstitial reduce the volumes of the interstitial and of the surrounding atoms in a characteristic way when the interstitial is in an otherwise perfect crystal;
Fig.\ \ref{fig:atomic_vol} shows an example of this phenomenon for BCC Fe.
The distribution of atomic volumes in our simulations is decomposed with a K-means algorithm into a superposition of peaks, one for each characteristic atomic volume surrounding an isolated interstitial.
Part of the utility of K-means is that the boundaries that define the locations the peaks can be dynamically updated from one time step to the next; this is necessary to account for the shifting and broadening of the peaks as the structure reaches higher defect concentrations.
Comparing the numbers of atoms assigned to each peak with the corresponding numbers of atoms for an isolated interstitial gives an estimate for the number of interstitials in our simulations, though the accuracy is expected to decrease as the peaks broaden and overlap with increasing radiation damage.

Stacking fault (SF) densities in the FCC CrCoNi and A-atom systems are relatively simple to evaluate by comparison.
Since the local atomic structure around atoms belonging to a SF appears to be HCP, all that is necessary for a reasonable estimate is to count the number atoms classified as HPC by, e.g., OVITO's common neighbor analysis, and to convert this to an equivalent SF area using the geometry of the crystal structure.  

The Warren-Cowley parameters are used to quantify the type and degree of chemical short range order (SRO), and are defined as \cite{cowley1950approximate,walsh2021magnetically}
\begin{equation}
\alpha_{ij} = 1 - p_{ij} / c_j
\label{eq:WC}
\end{equation}
where $c_j$ is the overall fraction of atomic type $j$, and $p_{ij}$ is the probability that an atom in the first nearest-neighbor shell of an atom with type $i$ is of type $j$.
Positive values of $\alpha_{ij}$ means there is a repulsion between species $i$ and $j$ while negative values indicate attraction.

\subsection{Energy model}
\label{subsec:energy_model}

This section develops a model to partition the energetic contributions of the various types of material defects to the overall potential energy change as a function of dpa.
The potential energy stored in defects is defined as $E_{\mathrm{stored}} = E_{\mathrm{pe}} - E_{\mathrm{coh}} - E_{\mathrm{elas}}$, where $E_{\mathrm{pe}}$ is the overall potential energy, $E_{\mathrm{coh}}$ is the cohesive energy of a reference material, and $E_{\mathrm{elas}}$ is the elastic strain energy.
This stored energy is modeled as a sum over defect contributions:
\begin{equation}
E_{\mathrm{fit}} = E_{\mathrm{dis}} + E_{\mathrm{vcy}} + E_{\mathrm{int}} - E_{\mathrm{sro}}
\label{eq:E_store}
\end{equation}
which includes terms for the contributions of dislocations, vacancies, interstitials, and chemical short range order, respectively, and where the sign of $E_\mathrm{sro}$ reflects the fact that increasing SRO decreases the system energy.
The contribution of stacking faults to $E_{\mathrm{fit}}$ was found to be negligible.

The energy of a well-developed dislocation network in an elastically isotropic material is well-described by the equation \cite{bertin2018}:
\begin{equation}
E_{\mathrm{dis}} = \chi \frac{\mu b^2}{4 \pi} \rho \ln \left(\frac{1}{r_c \sqrt{\rho}}\right)
\label{eq:E_dis}
\end{equation}
where $\mu$ is the shear modulus, $b$ is the Burgers vector, $r_c$ is a cutoff radius of the dislocation core stress field, and $\rho$ is the dislocation density.
The value of $r_c$ is approximated as the lattice parameter $a_0$.
$\chi$ is a parameter that accounts for the overall dislocation character, and depending on the material and the development of the network is expected to be in the interval $1.0 \leq \chi \leq 1/(1 - \nu)$ where $\nu$ is Poisson's ratio.
The elastic constants that are necessary to evaluate Eq.\ \ref{eq:E_dis} were found in Refs.\ \cite{jian2020effects, mendelev2003development}, and the material parameters used in the model are included in Table \ref{table:Dis_param}.

\begin{table}
    \caption{Material properties used as parameters to evaluate the dislocation energy.
    The values for the CrCoNi and A-atom materials can be found in Refs.\ \cite{jian2020effects, mendelev2003development}.}
    \begin{center}
        \begin{tabular}{@{}ccccc@{}} \toprule
        & $\mu \ [\mathrm{GPa}]$ &\ $b \ [\mathrm{A}]$ &\ $a_0\ [\mathrm{A}]$&\ $\nu$ \\
        \midrule
        Fe & $49.5$ & $2.49$ & $2.87$ & $0.373$ \\
        CrCoNi & $37.0$ & $2.47$ & $3.50$ & $0.414$ \\
        A-atom & $43.0$ & $2.47$ & $3.50$ & $0.403$ \\
        \bottomrule
        \end{tabular} \label{table:Dis_param}
    \end{center}
\end{table}

The energetic contributions of the point defects are modeled as: 
\begin{align}
E_{\mathrm{vcy}} &= \beta e_{\mathrm{vcy}} N_{\mathrm{vcy}}
\label{eq:E_vcy} \\
E_{\mathrm{int}} &= \gamma e_{\mathrm{int}} N_{\mathrm{int}}
\label{eq:E_int}
\end{align}
where $e_{\mathrm{vcy}}$ and $e_{\mathrm{int}}$ are the formation energies of an isolated vacancy or interstitial as found by inserting a single vacancy or interstitial into an otherwise perfect crystal.
More specifically, after minimizing the potential energy and measuring the potential energy change $\Delta E$ relative to the perfect crystal, the isolated point defect formation energies are defined as: 
\begin{align}
e_{\mathrm{vcy}} &= \Delta E_{\mathrm{remove}} + E_{\mathrm{coh}} 
\label{e_vcy} \\
e_{\mathrm{int}} &= \Delta E_{\mathrm{add}} - E_{\mathrm{coh}}.
\label{e_int}
\end{align}
The values of these formation energies are reported in Table \ref{table:point_defect}.
$N_{\mathrm{vcy}}$ and $N_{\mathrm{int}}$ are the numbers of vacancies and interstitials, and
$\beta$ and $\gamma$ are fitting parameters to account for deviations from the isolated point defect energies that occur as the density of point defects increases with dpa.

\begin{table}
    \caption{Isolated point defect formation energies.} 
    \begin{center}
        \begin{tabular}{@{}ccc@{}} \toprule
        & $e_{\mathrm{vcy}}\ [\mathrm{eV}]$ &\ $e_{\mathrm{int}} \ [\mathrm{eV}]$ \\
        \midrule
        Fe & $1.71$ & $4.01$ \\
        CrCoNi & $1.49$ & $2.45$\\
        A-atom & $1.62$ & $3.85$\\
        \bottomrule
        \end{tabular} \label{table:point_defect}
    \end{center} 
\end{table}

Finally, $E_{\mathrm{sro}}$ is the energy associated with the chemical short range order relative to a random solid solution, and was evaluated for the CrCoNi MPEA system using molecular statics calculations.
Starting with an initial configuration, the types of all the atoms were randomly reassigned to one of the three constituent elements and the potential energy of the structure was minimized.
This procedure was repeated five times for each configuration to account for statistical variations, and the the average potential energy change relative to the initial structure was included in the model for each dpa for which the fitting was conducted.
Overall, the energy contribution of stacking faults is considered negligible, with an average value over all dpa of only $-0.0463\ \mathrm{meV/atom}$ for the CrCoNi system.

The resulting energy model contains the three adjustable parameters $\chi$, $\beta$, and $\gamma$.
Fitting the model to each of the three systems involved first constructing the bounds on the dislocation energy that would be realized by the minimum and maximum allowed values for $\chi$.
The dislocation energy was then fixed at the lower bound, and a least-squares technique was used to find the values of $\beta$ and $\gamma$ that minimized the difference between $E_{\mathrm{stored}}$ and $E_{\mathrm{fit}}$ over the entire interval up to $2.0$ dpa.
Repeating this procedure with the dislocation energy fixed at the upper bound allowed the construction of corresponding intervals for the predicted contributions of the vacancy and interstitial terms.
As discussed below in Sec.\ \ref{subsec:point_defects}, the largest contribution to the model error is likely the estimated numbers of point defects $N_{\mathrm{vcy}}$ and $N_{\mathrm{int}}$.

\section{Results and Discussion}
\label{sec:results_and_Discussion}




\subsection{Total energy and pressure}
\label{subsec:total_energy}

\begin{figure}
	\centering
\includegraphics[width=0.40\textwidth]{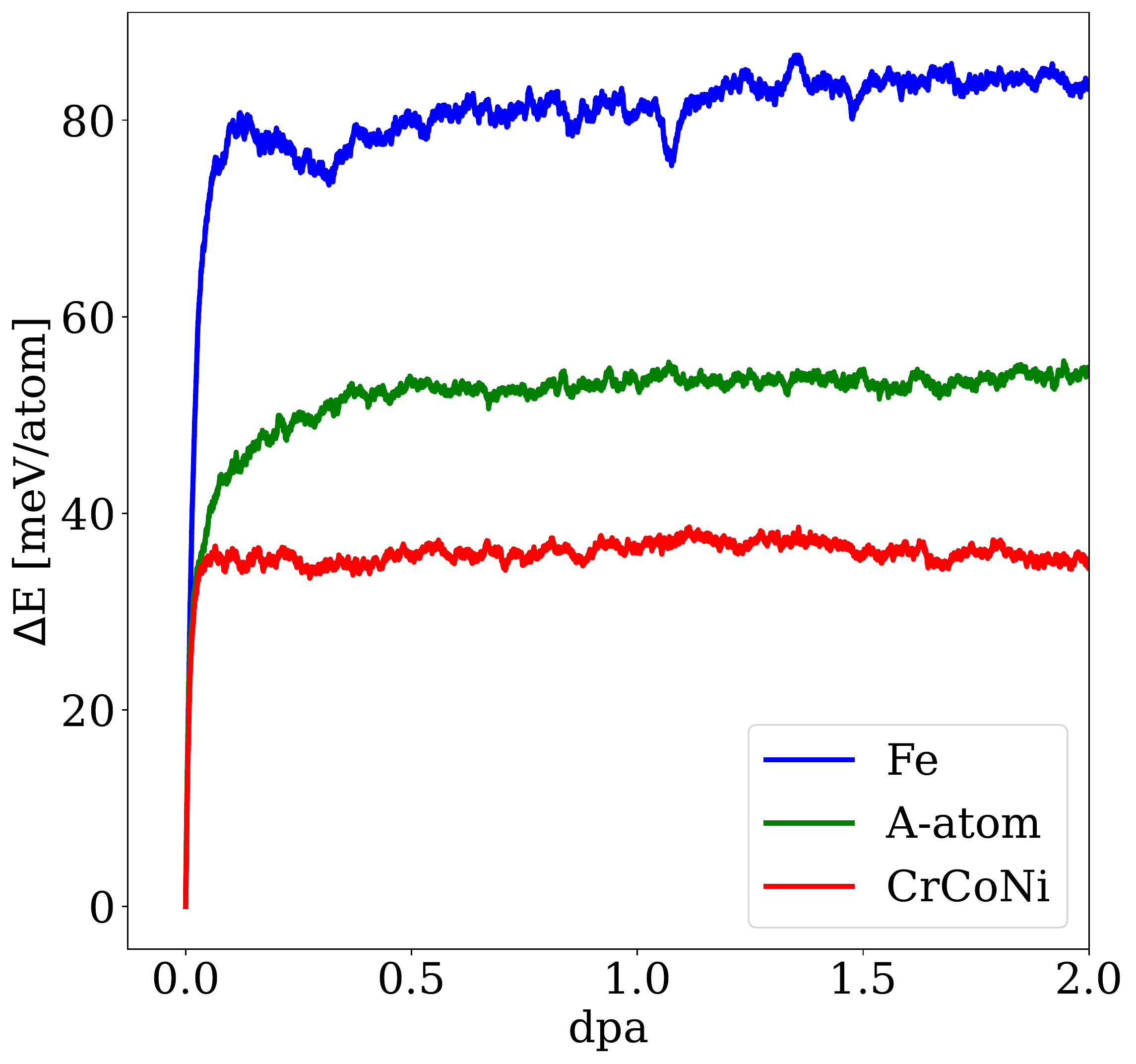}
	\caption{The change in the potential energy of the simulation cell as a function of dpa for all three material systems.}
	\label{fig:all_delta_e_vs_dpa}
\end{figure}

\begin{figure}
	\centering
    {\includegraphics[width=0.40\textwidth]{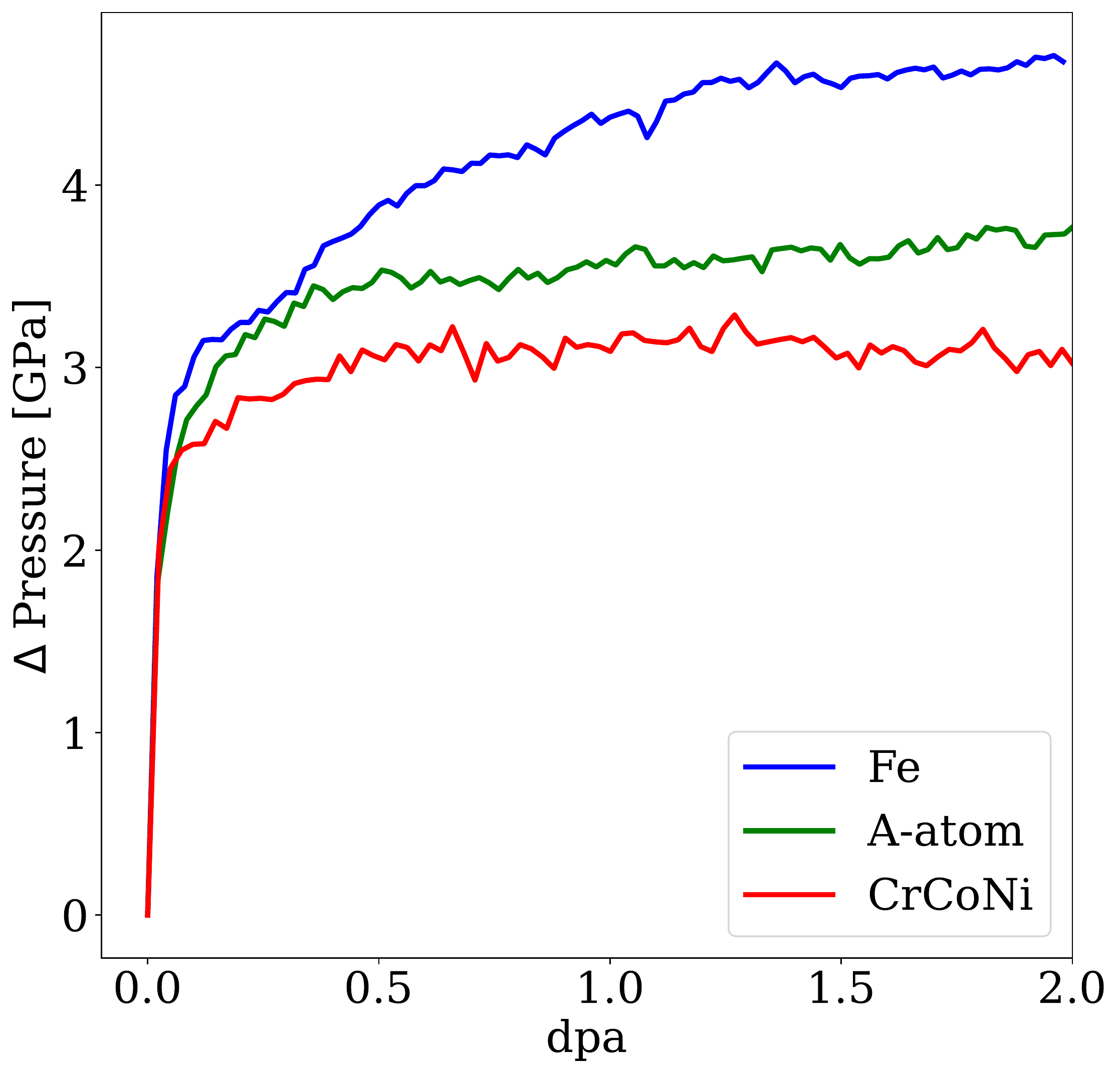}}
	\caption{The pressure of the simulation cell as a function of dpa for all three material systems.}
	\label{fig:all_pres_vs_dpa}
\end{figure}

The successive generation of Frenkel pairs considerably increases the potential energy of the simulated systems, particularly in the absence of thermally-driven point defect migration and recombination.
As displayed in Fig.\ \ref{fig:all_delta_e_vs_dpa}, the energies of all three systems roughly reach steady states by $0.5$ dpa, suggesting the activation of a recovery mechanism that offsets the energy increase of additional Frenkel pairs.
The convergence of the potential energy does not indicate that the microstructure has reached a steady state though;
a slow but continual increase in pressure past $0.5$ dpa that is most visible for the Fe system in Fig.\ \ref{fig:all_pres_vs_dpa} implies that at least some features of the microstructure continue to evolve up to much higher dpa.

The supplied energy that persists through the structural relaxations can be partitioned into local material defects and an overall elastic energy---there is no kinetic energy in the absence of atomic velocities.
The increasing elastic energy is a direct consequence of the introduction of Frenkel pairs; consider that displacing an internal atom to an external surface of a crystal increases the crystal volume by one atomic volume.
The simulations are performed at fixed volume though, meaning that the boundary conditions can only be satisfied by subjecting the crystal to an increasing pressure to maintain a net zero volumetric strain.
The elastic energy is defined as the elastic work that would need to be performed on the relaxed system in a zero pressure configuration to return it to the required volume.
This can be calculated from the material's elastic constants and the volume of the system in the relaxed configuration, with the later evaluated by allowing the simulation box to expand while minimizing the potential energy.
The percent of the supplied energy that resides as elastic energy is negligible in all situations, being $3\%$, $0.2\%$ and $0.3\%$ for the Fe, CrCoNi and A-atom systems, respectively.
The overwhelming majority of the supplied energy therefore resides in the form of defects, specifically dislocations, interstitials, and vacancies (the contribution of stacking faults was found to be negligible in Sec.\ \ref{subsec:energy_model}).

\subsection{Dislocations}
\label{subsec:dislocations}

\begin{figure}
	\centering
    {\includegraphics[width=0.25\textwidth]{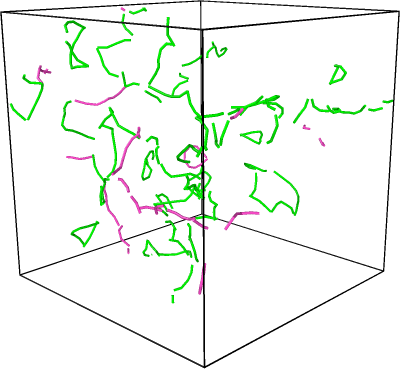}} \
	{\includegraphics[width=0.25\textwidth]{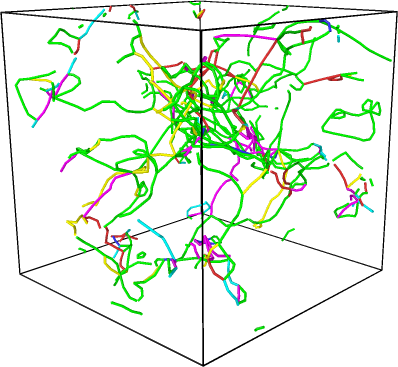}} \
	{\includegraphics[width=0.25\textwidth]{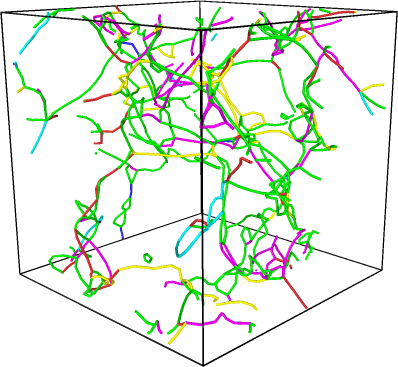}}

	\vspace{0.1cm}
	{\includegraphics[width=0.25\textwidth]{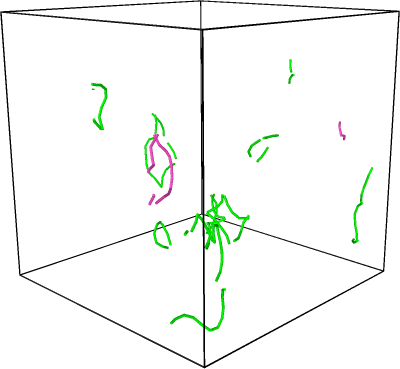}} \
	{\includegraphics[width=0.25\textwidth]{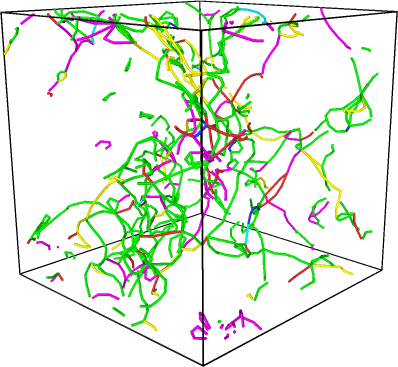}} \
	{\includegraphics[width=0.25\textwidth]{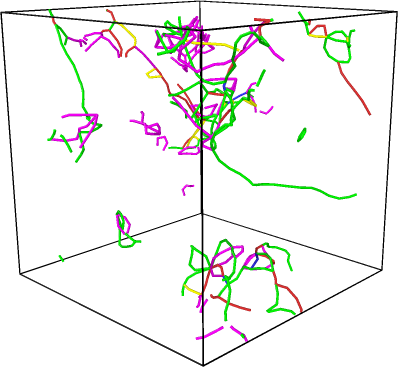}}
	\caption{Dislocation networks at $0.5$ dpa (top) and $2.0$ dpa (bottom) for Fe (left), CrCoNi (middle), and A-atom (right) systems where color indicates dislocation types.
	For the Fe system green are $1/2 \langle 111 \rangle$ and purple are $\langle 100 \rangle$ dislocations.
	For the CrCoNi and A-atom systems green are $1/6\left<112\right>$ Shockley partials, purple are $1/6\left<110\right>$ stair-rod, yellow are $1/3\left<100\right>$ Hirth, light blue are $1/3\left<111\right>$ Frank, and dark blue are $1/2\left<110\right>$ perfect dislocations.}
	\label{fig:Dislocations_0.5dpa}
\end{figure}

\begin{figure}
	\centering
    {\includegraphics[width=0.32\textwidth]{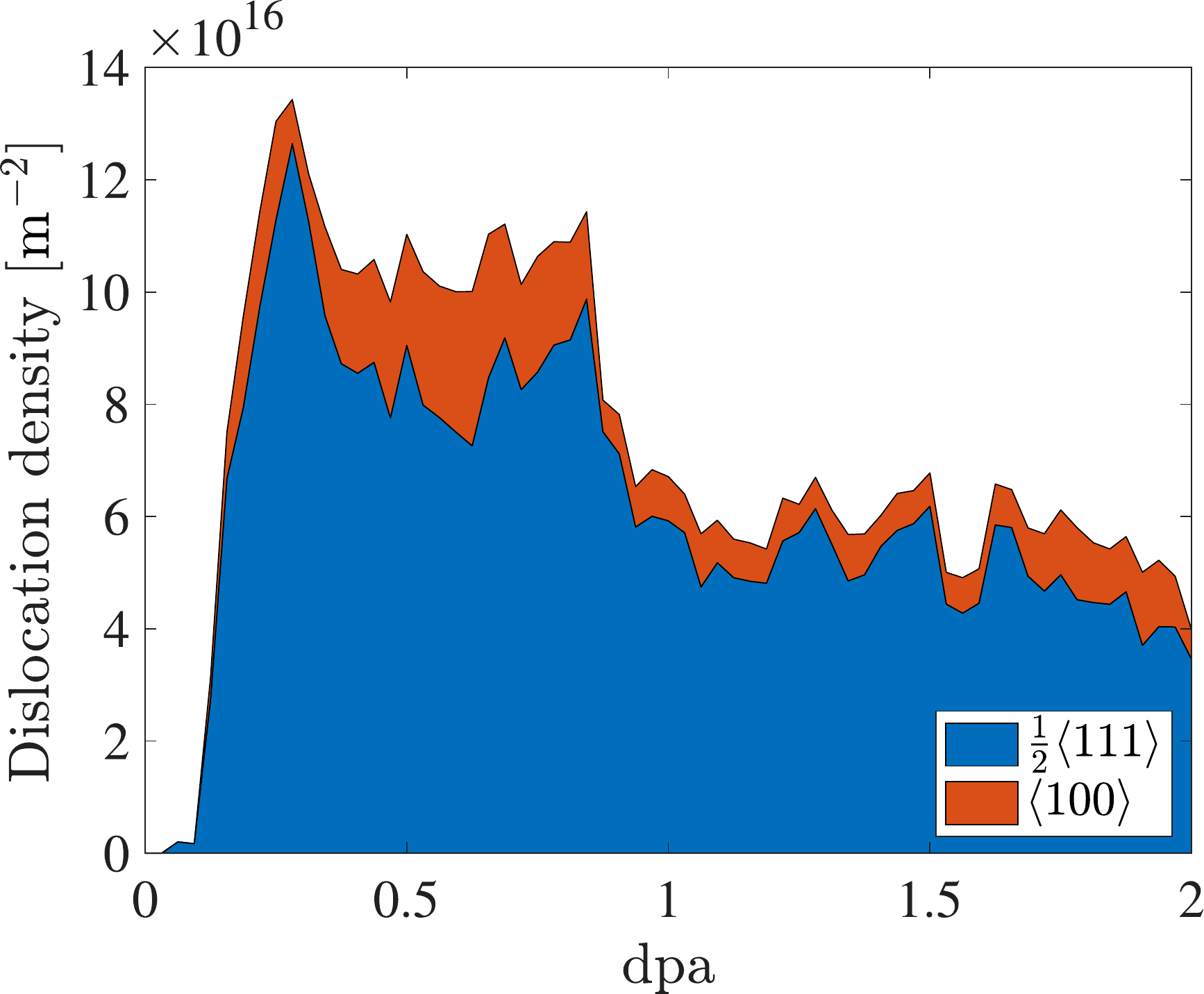}}
	{\includegraphics[width=0.32\textwidth]{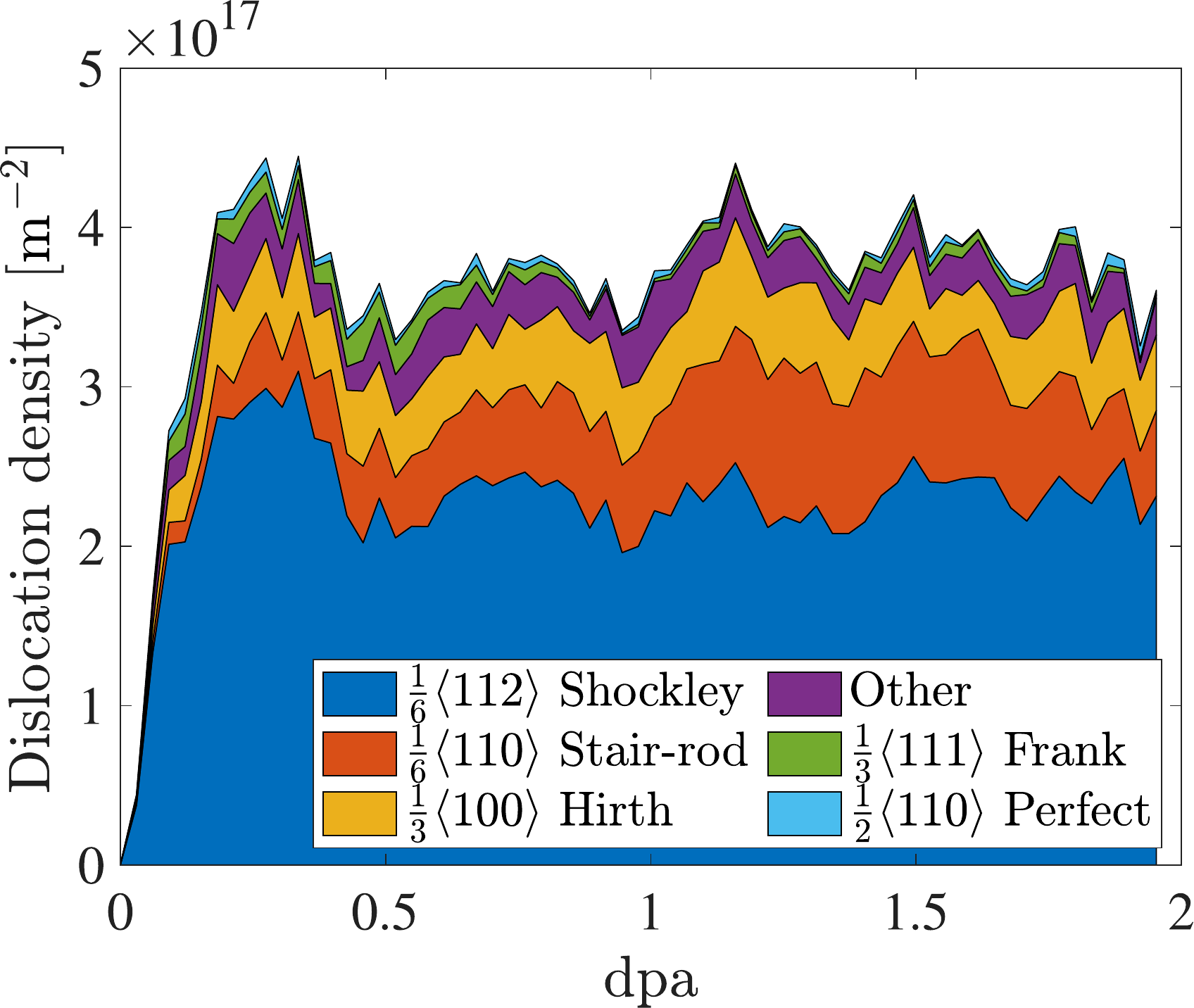}}
	{\includegraphics[width=0.32\textwidth]{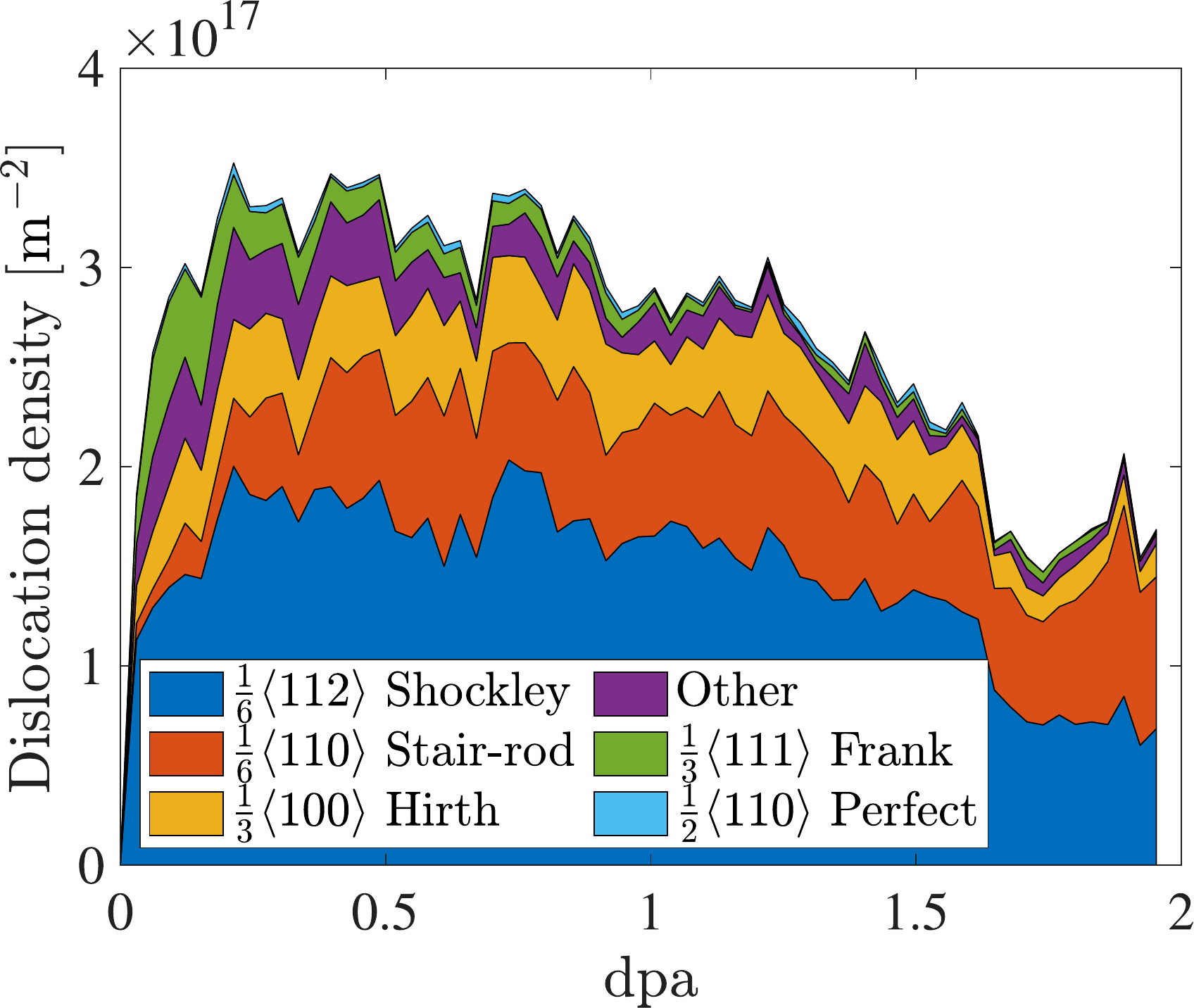}}
	\caption{The dislocation density and dislocation types for Fe (left), CrCoNi (middle), and A-atom (right) as a function of dpa.}
	\label{fig:all_dd}
\end{figure}

While the dislocation networks that developed in the three systems are clearly distinct, the dislocation densities of all three initially rapidly increased before falling back to steady state values.
Figure \ref{fig:Dislocations_0.5dpa} shows the dislocation networks in the Fe (left), CrCoNi (middle), and A-atom (right) systems at $0.5$ dpa (top) and $2.0$ dpa (bottom) as found by the DXA, with the network density visibly lower at higher dpa for the Fe and A-atom systems.
A more quantitative analysis of the distribution of dislocation types as a function of dpa is provided in Fig.\ \ref{fig:all_dd} where the dislocation density for the Fe system is visibly lower than that for both the FCC systems at all dpa.
This can be explained by the main dislocation production mechanism in irradiated materials; isolated self-interstitials precipitate as interstitial disks, forming dislocations loops that evolve and eventually develop into a larger network.
In Fe this is known to produce more mobile $1/2 \langle 111 \rangle$ and less mobile $\langle 100 \rangle$ loops, with the population of the former being greater at low temperatures \cite{marian2002mechanism,arakawa2011direct}.
The $1/2 \langle 111 \rangle$ dislocations begin as isolated loops, but appear to be mobile enough to migrate and react once the dislocation density and internal stresses reach critical values around $0.8$ dpa, reducing the overall dislocation density as a dislocation network is formed.


\begin{figure}
	\centering
    {\includegraphics[width=0.25\textwidth]{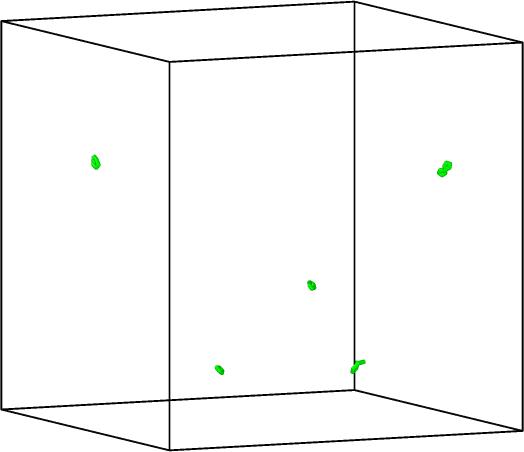}}
	{\includegraphics[width=0.25\textwidth]{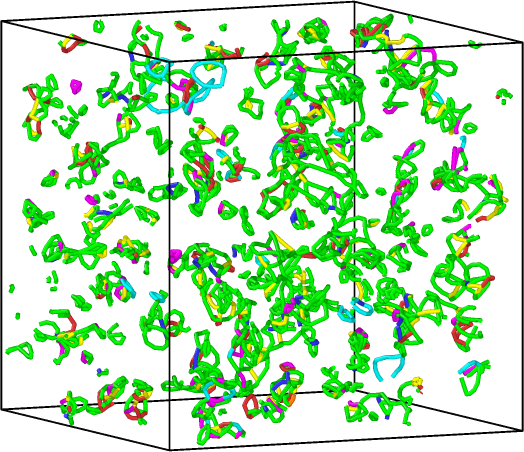}}
	{\includegraphics[width=0.25\textwidth]{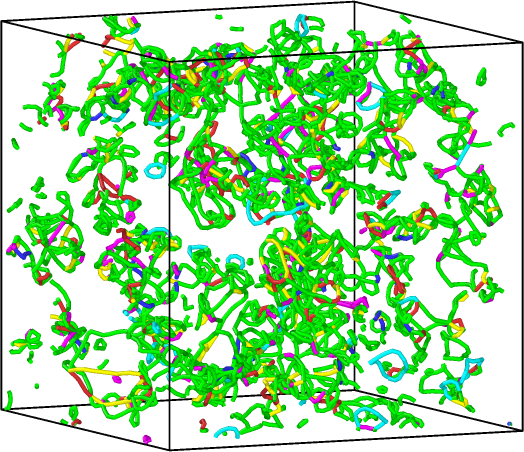}}
	\caption{The dislocation network for a CrCoNi system containing approximately one million atoms at $0.007$ (left), $0.01$ (middle) and $0.02$ (right) dpa. Green are $1/6\left<112\right>$ Shockley partials, purple are $1/6\left<110\right>$ stair-rods, yellow are $1/3\left<100\right>$ Hirth, light blue are $1/3\left<111\right>$ Frank, and dark blue are $1/2\left<110\right>$ perfect dislocations.}
	\label{fig:dis_den_CrCoNi_MAR}
\end{figure}

The expectation was that the dislocation networks in the CrCoNi and A-atom systems would begin as isolated extrinsic Frank loops resulting from the precipitation of interstitials on $\{111\}$ planes.
It is well established experimentally that such sessile Frank loops eventually unfault to form a glissle dislocation network containing both perfect dislocations and Shockley partials, though the precise mechanism by which this occurs continues to be a subject of study \cite{gelles1981frank,rodney2004molecular,kadoyoshi2007molecular}.
What was unexpected in Fig.\ \ref{fig:all_dd} is that this unfaulting process should apparently begin at the outset, with the population of Shockley partials visibly exceeding that of any other dislocation type in Fig.\ \ref{fig:all_dd} even for very low dpa.
A similar behavior is observed for the larger CrCoNi simulation described in Appendix \ref{sec:appendix}, indicating that this is not merely a finite size effect, and Fig.\ \ref{fig:dis_den_CrCoNi_MAR} even suggests that Shockley partial loops could be nucleating directly at the earliest stages of radiation damage.
While there is a measurable population of extrinsic Frank loops that slowly decreases up to around $1.0$ dpa, the rapid growth of the population of Shockley partials well before this point strongly suggests that there is some other mechanism by which Shockley partials are being generated.
Understanding the details of the dislocation formation mechanisms at the earliest stages of radiation damage would require a dedicated study that is relegated to future work though.

The generally increasing populations of sessile stair-rod dislocations with dpa in the CrCoNi and A-atom systems is an expected result of the maturation of a glissile dislocation network mainly composed of Shockely partials.
Finally, it is significant that the dislocation network in the A-atom system continues to undergo substantial change even up to $1.6$ dpa, well beyond the $0.5$ dpa at which the potential energy converged in Fig.\ \ref{fig:all_delta_e_vs_dpa}.
The invariance of the potential energy even as the dislocation density is reduced by nearly half requires a corresponding increase in the population of other defect types, a point that will be significant in the following.



\subsection{Stacking faults}
\label{subsec:stacking_faults}

\begin{figure}
	\centering
    {\includegraphics[width=0.24\textwidth]{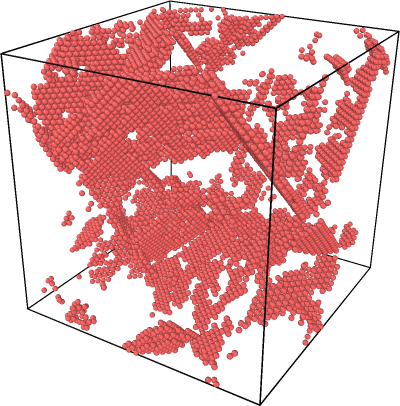}}
	{\includegraphics[width=0.24\textwidth]{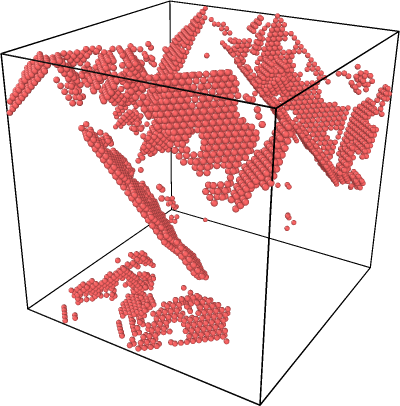}}
	\qquad
	\caption{Stacking faults as identified by locally HCP coordinated atoms in CrCoNi (left) and A-atom (right) systems at $2.0$ dpa.}
	\label{fig:SF_FCC}
\end{figure}

\begin{figure}
	\centering
    {\includegraphics[width=0.45\textwidth]{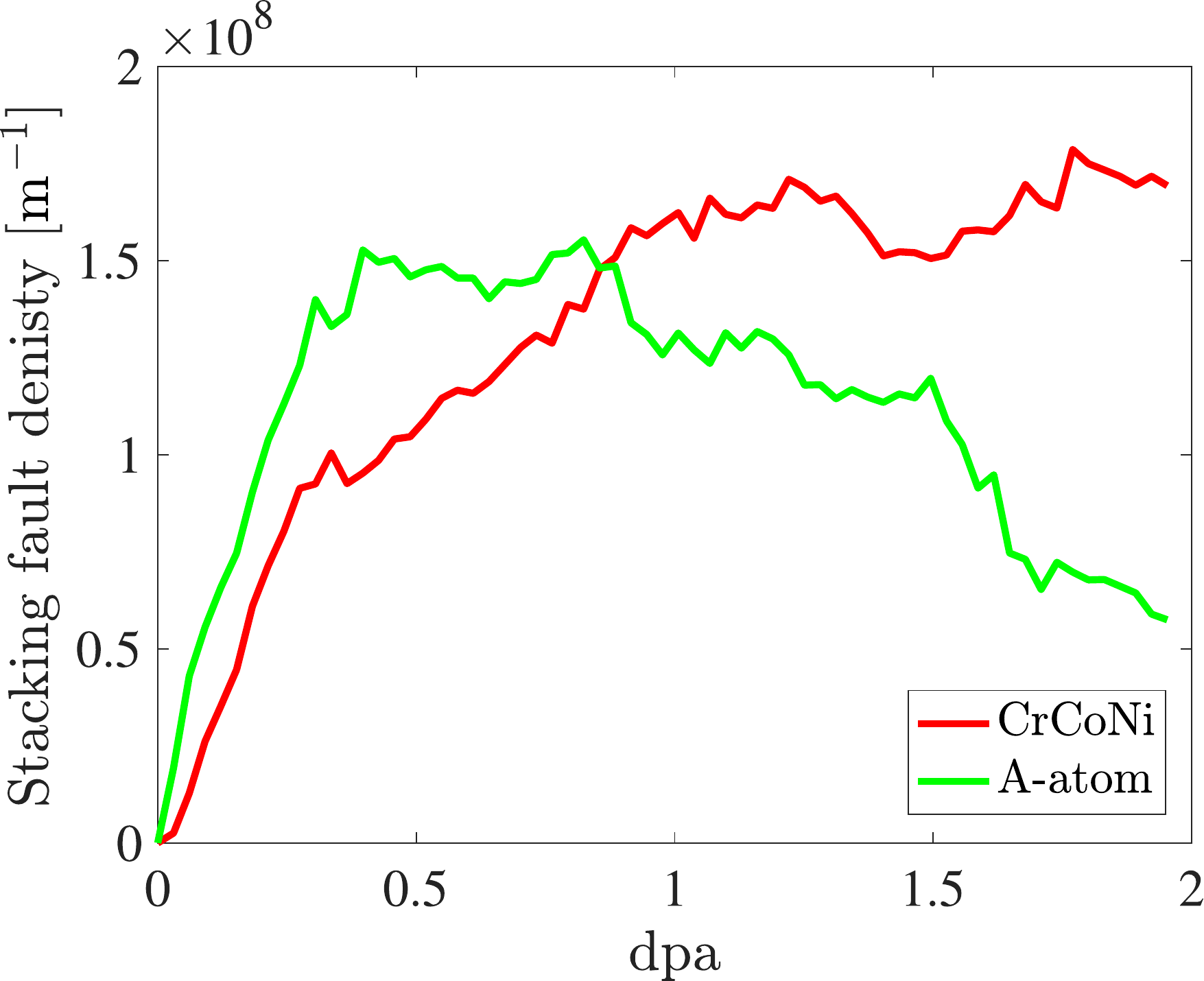}}
	\caption{The stacking fault density as a function of dpa for the CrCoNi and A-atom systems.}
	\label{fig:Stacking_fault_density}
\end{figure}

As expected, stacking faults (SFs) were only observed in the FCC CrCoNi and A-atom systems.
The SF configurations at $2.0$ dpa are shown in Fig.\ \ref{fig:SF_FCC}, and the SF density as a function of dpa is reported in Fig.\ \ref{fig:Stacking_fault_density}.
The very high SF densities observed in these systems are consistent with the low reported values for their SF energies \cite{jian2020effects} and with the high density of partial dislocations in Fig.\ \ref{fig:all_dd}.
It is interesting that the SF density for the A-atom system was higher than for the CrCoNi system up to around $0.8$ dpa, but by $2.0$ dpa the situation had reversed with the SF density in the A-atom system falling by almost a factor of three.
This roughly correlates with the decrease in the density of Shockley partials in Fig.\ \ref{fig:all_dd}, though the correspondence is not exact.
If the decrease in the SF density is in fact driven by the maturation of the dislocation network, then an important question is why a similar decrease in the density of Shockley partials and SFs was not observed in the CrCoNi system.
Any explanation should likely involve the chemical SRO since this is (by design) the main difference between the CrCoNi and A-atom systems.
Possible mechanisms include atomic rearrangements following the formation of the SF lowering the energy of the SF relative to the pristine state and increasing the unfaulting barrier, or the fluctuations in the local stacking fault energy decreasing the difficulty of cross slip and increasing the complexity of the dislocation network.
Either of these mechanisms would help to explain why the dislocation density for the CrCoNi system does not fall as quickly with increasing dpa as for the A-atom system.

The elevated SF density observed here could have significant implications for the mechanical strength of irradiated CrCoNi since SFs can act as barriers to dislocation motion that increase plastic strength.
Recent work by Richie et al.\ used \textit{in situ} transmission electron microscopy to investigate CrCoNi under cryogenic conditions and noted that a high density of SFs and extensive cross slip likely contributed to the superior mechanical properties of the alloy \cite{ding2019real}. 
SFs have also been suggested to decrease radiation-induced void swelling of CrCoNi by alleviating internal stress build up \cite{lu2019irradiation}.

\subsection{Short range ordering}
\label{subsec_sro}

\begin{figure}
	\centering
	\includegraphics[width=0.45\textwidth]{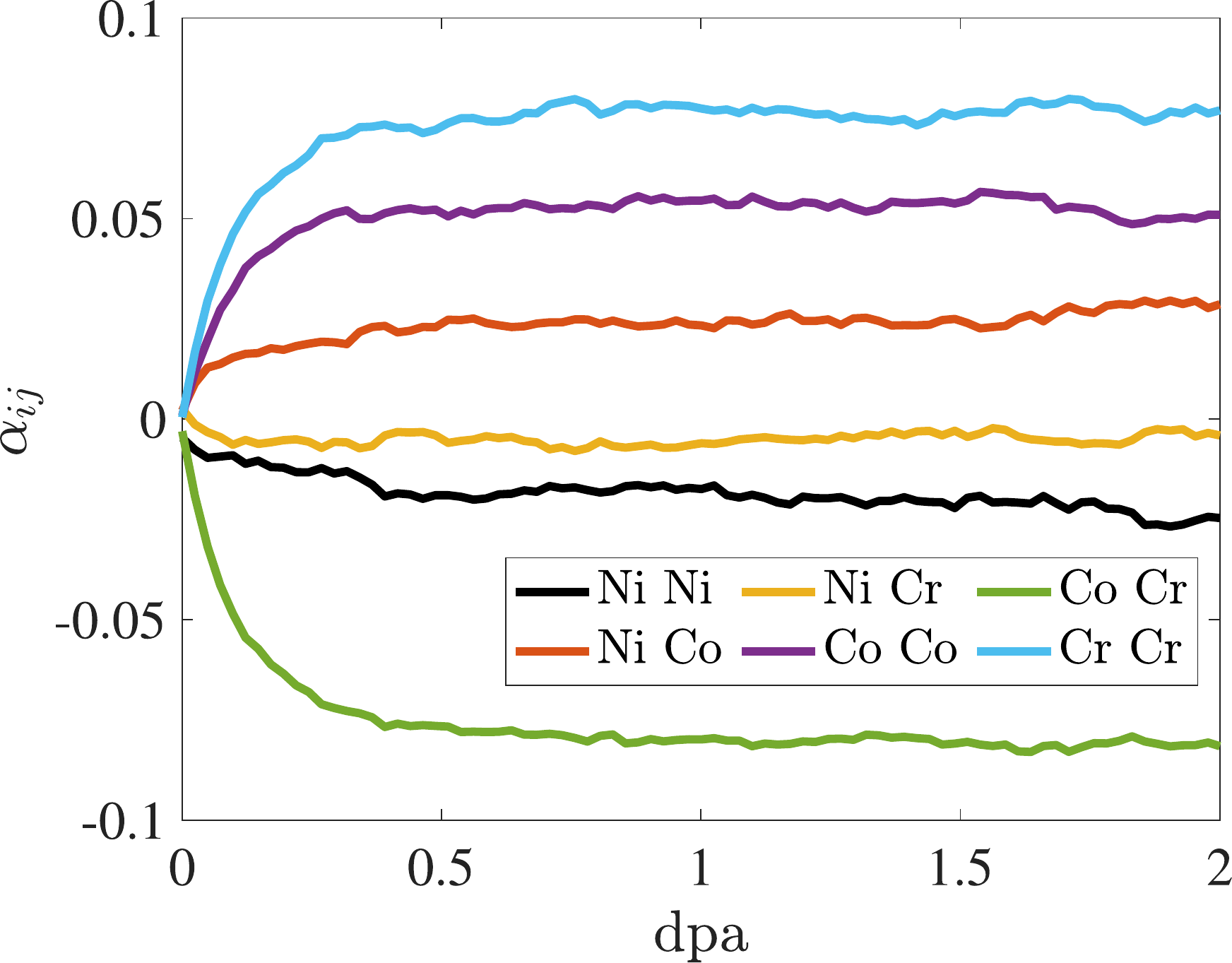}
	\caption{The Warren-Cowley SRO parameters as a function of dpa for the CrCoNi system.}
	\label{fig:WC}
\end{figure}

The CrCoNi system, initialized as a random solid solution, did develop a minor degree of SRO during the CRA simulation.
While not visually apparent this ordering was statistically significant as indicated by the WC SRO parameters reported in Fig.\ \ref{fig:WC}.
All of the WC parameters stabilized by $0.5$ dpa, with Cr displaying an increased likelihood to neighbor Co and both Cr and Co less likely to neighbor other atoms of the same species.
It is interesting that this degree of order developed despite the randomizing effects of atomic displacements in the CRA, and is significant that the order generally agrees with what has been found experimentally \cite{zhang2017local} and with more accurate DFT-based Monte Carlo simulations \cite{2018ding}.
Finally, despite the magnitude of the SRO being relatively small, the contribution toward lowering the potential energy of the system is still substantial;
as will be discussed in Sec.\ \ref{subsec:point_defects} below, the magnitude of the $E_{\mathrm{sro}}$ can be as much as $30\%$ of the magnitude of $E_{\mathrm{fit}}$.

\subsection{Point defects and energy balance}
\label{subsec:point_defects}

The results discussed up to this point have included dislocations and stacking faults, i.e., defects for which there are standard identification techniques and about which we have general confidence.
The same is not true for point defects, for the identification of which there are arguably no methods in the literature (including the ones used here) that work reliably at high defect concentrations.
This section not only reports the nominal point defect concentrations as calculated using the methods described in Sec.\ \ref{sec:methods}, but highlights the inconsistency of these estimates with respect to the stored energy.
This is particularly concerning since the point defect balance equations that describe the evolution of point defect concentrations are the foundations of our understanding of damage development in irradiated materials \cite{mansur1994theory,was2016fundamentals}, and it is unclear whether it is currently possible to evaluate the independent variables in these equations either by experiment or simulation.
Our sincere hope is that this will be identified as an area requiring the further attention of the research community in the future. 

\begin{figure}
	\centering
    {\includegraphics[width=0.45\textwidth]{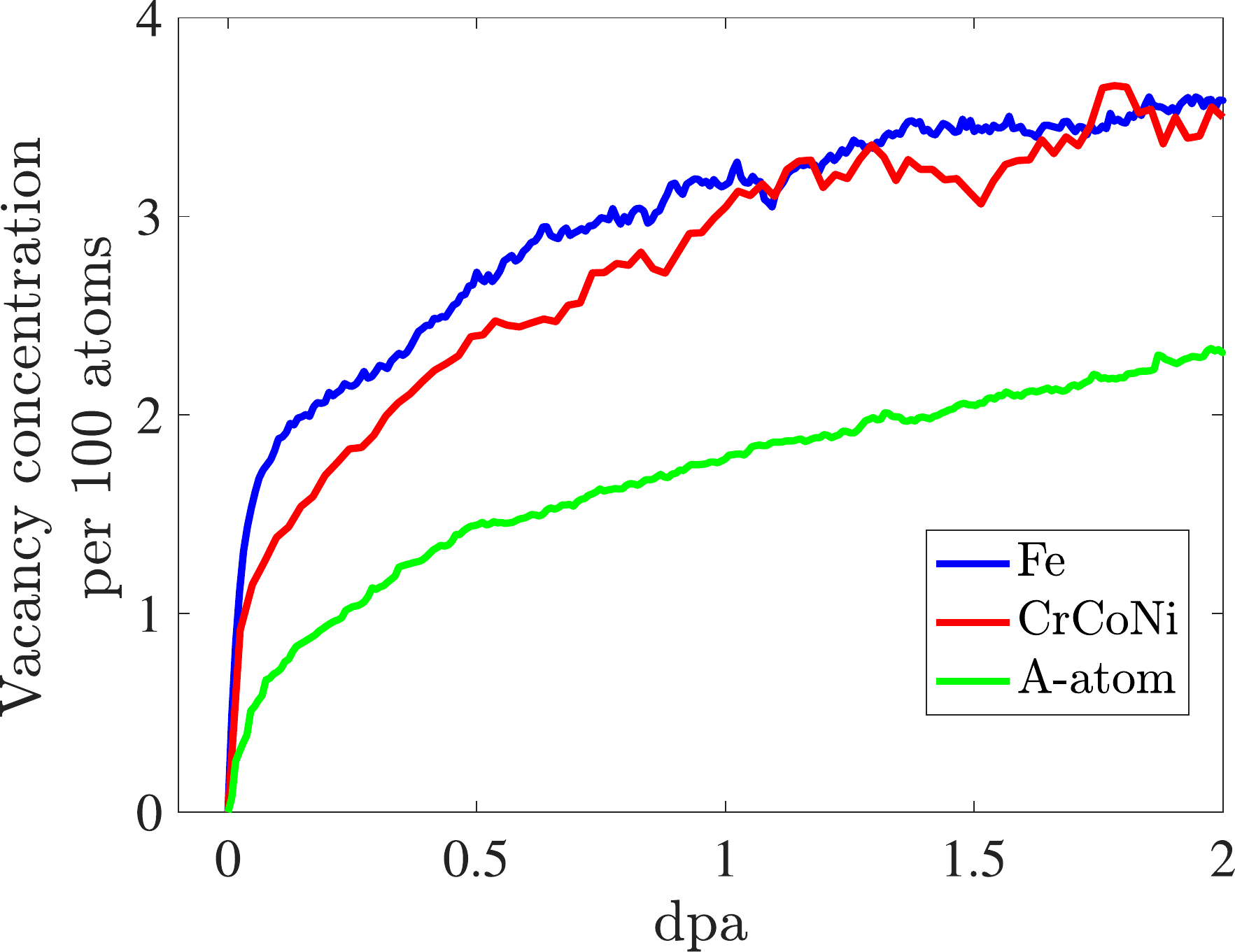}}\ \
	{\includegraphics[width=0.45\textwidth]{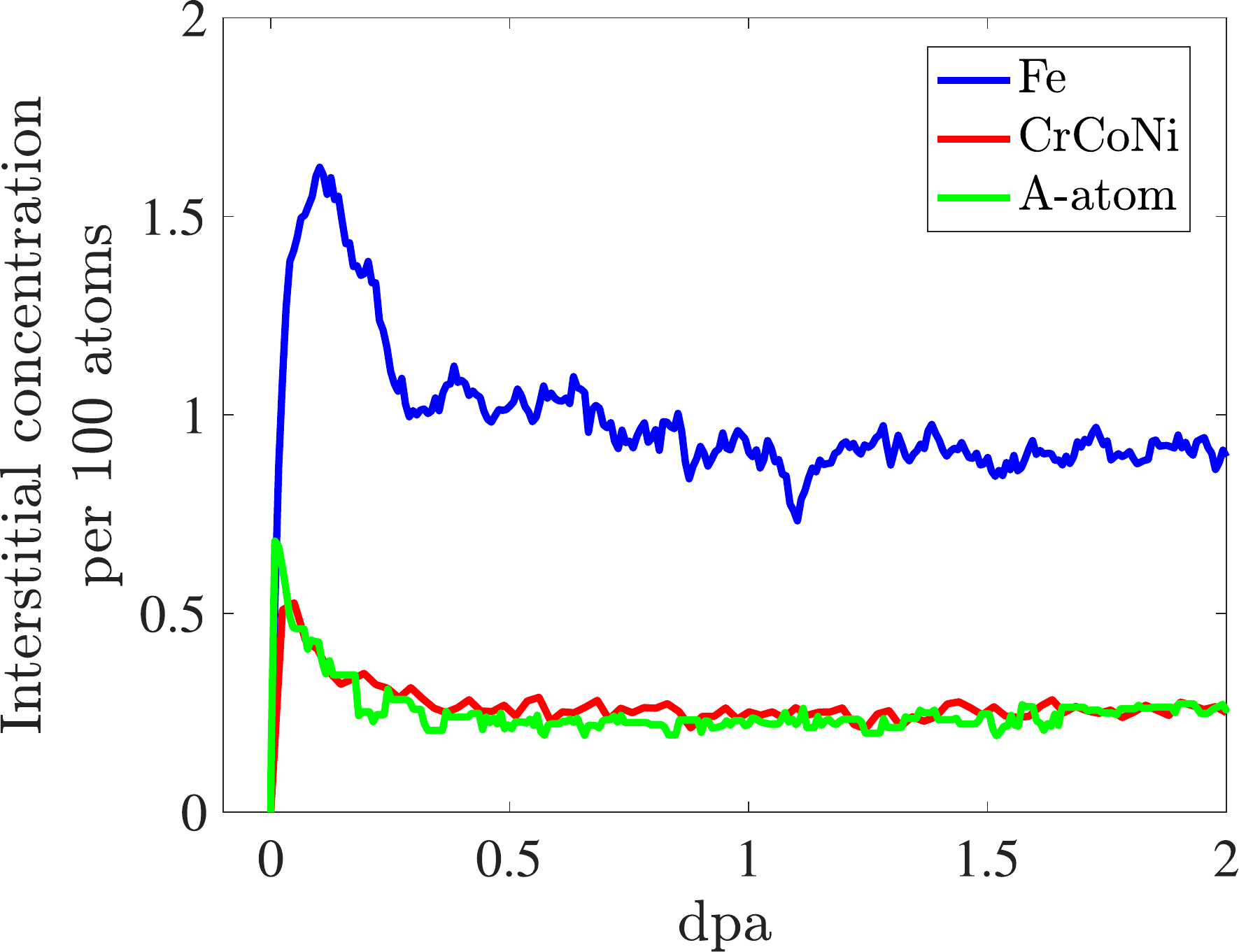}}
	\qquad
	\caption{Estimated vacancy and interstitial concentrations as measured by the WS and K-means methods, respectively, as a function of dpa for all three material systems.}
	\label{fig:Point_defects}
\end{figure}

One expectation is that the concentrations of vacancies and intertsitials should be equal at low dpa and increase linearly with the number of displacements.
This is because both are initially produced in equal amounts by Frenkel pair insertions and the defect density is not high enough for there to be appreciable recombination.
While this is satisfied for the point defect concentrations reported in Fig.\ \ref{fig:Point_defects}, the vacancy and interstitial concentrations quickly diverge with increasing dpa just as in the CRA simulations of Ref.\ \cite{Derlet2020}.
This is conventionally believed to indicate that the more mobile interstitials are clustering to form dislocation loops or are annihilating on previously-existing loops once a threshold density is reached, with the less mobile vacancies remaining in the bulk.
Although our simulations only considered single crystal systems, interstitials can also migrate to and annihilate on other types of sinks (e.g., grain boundaries, phase boundaries, voids) in more general materials.
The nominal vacancy concentrations of the Fe and CrCoNi systems were comparable at all dpa, and were consistently $\about 50\%$ greater than that of the A-atom system. 
It is suspicious that the vacancy concentrations of the two FCC systems as found by the WS method continue to increase all the way to $2.0$ dpa though, considering that the pressures of these systems have already converged by $0.5$ dpa in Fig.\ \ref{fig:all_pres_vs_dpa};
along with the results of the energy fitting reported below, this suggests that the WS method increasingly overestimates the vacancy concentrations of the FCC systems at higher dpa. 
Confusingly, the same does not seem to be true for the Fe system for which the vacancy concentration, pressure, and energy model results are all consistent.
The reason for this discrepancy is unknown, but highlights the need for more robust ways to identify vacancies (or a more precise definition of what constitutes a vacancy) in highly damaged structures.

\begin{figure}
	\centering
    {\includegraphics[width=0.32\textwidth]{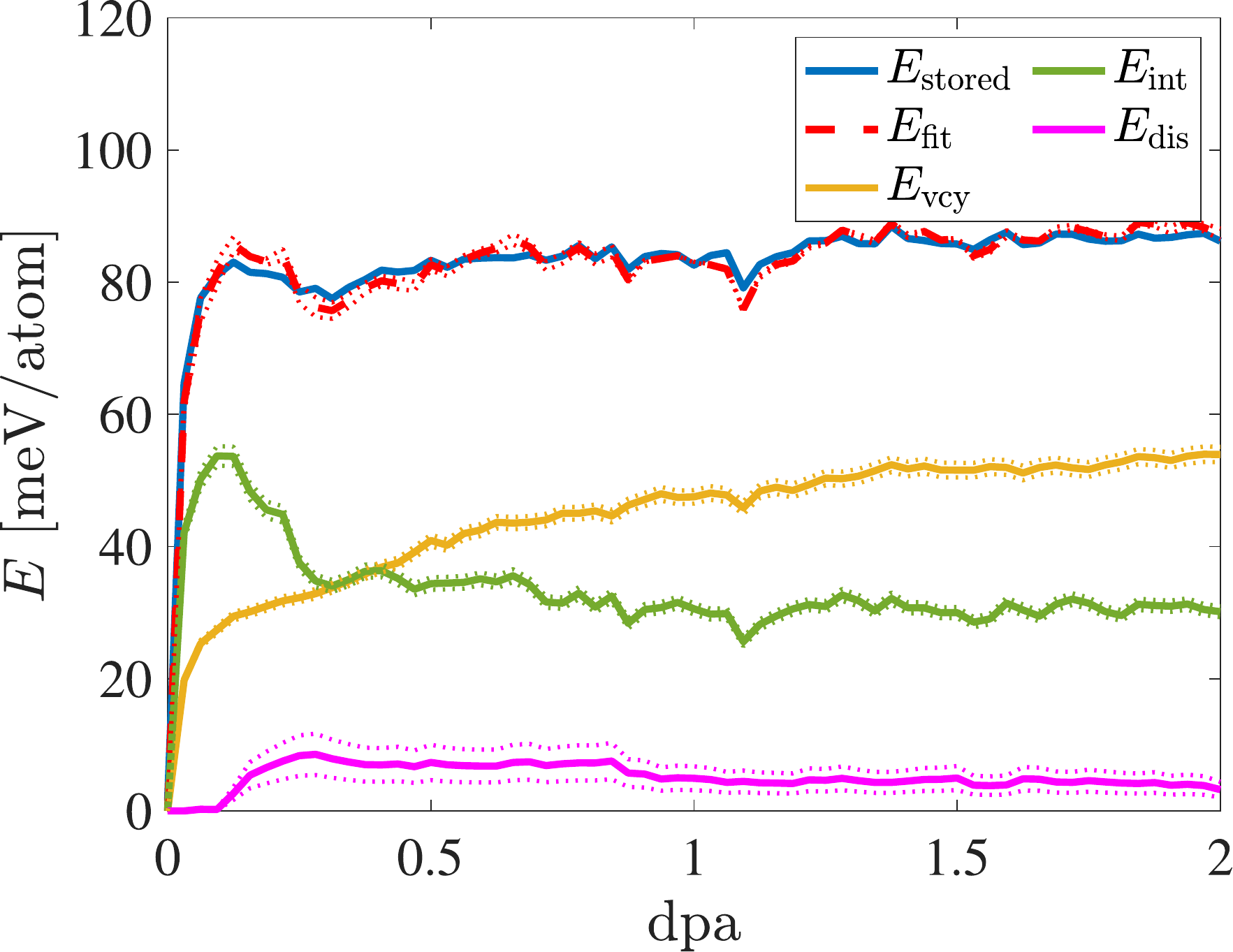}}
	{\includegraphics[width=0.32\textwidth]{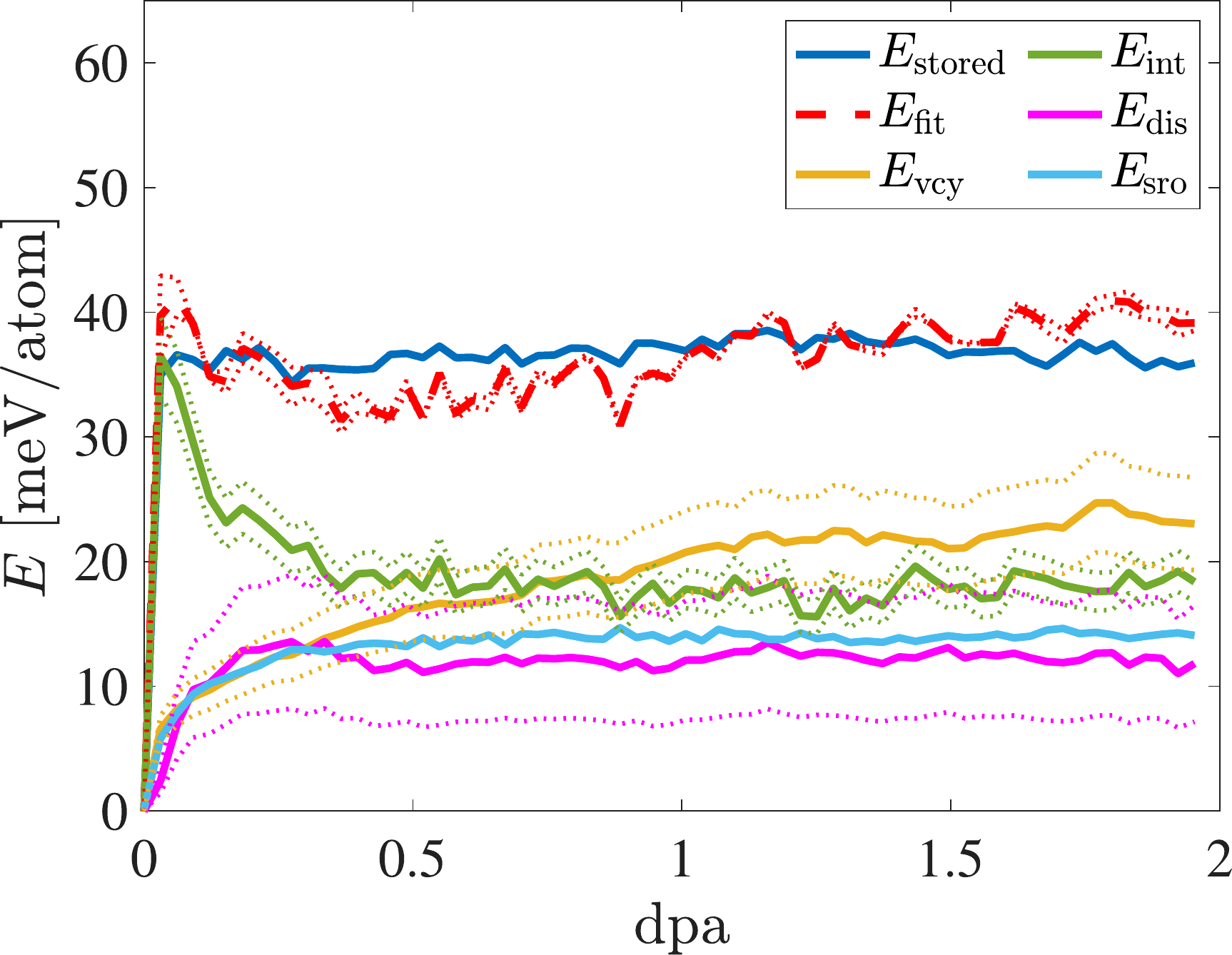}}
	{\includegraphics[width=0.32\textwidth]{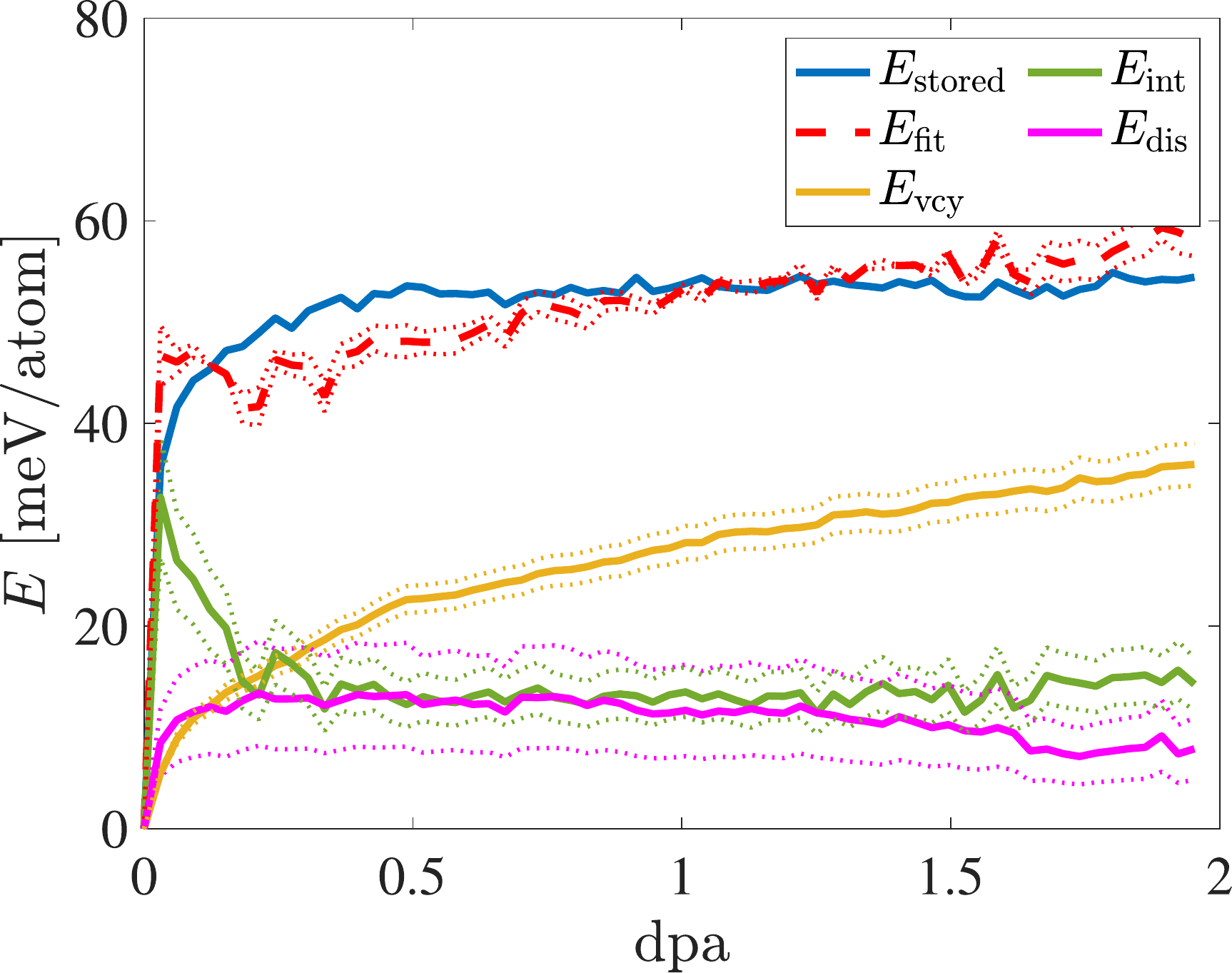}}
	\caption{Fit of the stored energy model described in Sec.\ \ref{subsec:energy_model} for the Fe (left), CrCoNi (center), and A-atom (right) systems.
	The dotted trend lines are bounds derived from the estimated bounds on the dislocation network energy.}
	\label{fig:Energy_fit}
\end{figure}

The results of fitting the energy model $E_\mathrm{fit}$ of Sec.\ \ref{subsec:energy_model} to $E_\mathrm{stored}$ for all three structures are reported in Fig.\ \ref{fig:Energy_fit}, with the values of the fitting parameters $\beta$ and $\gamma$ given in Table \ref{table:1}.
The model works remarkably well for the Fe system, with $E_\mathrm{fit}$ reproducing all of the general trends of $E_\mathrm{stored}$ and many of the smaller features as well.
The fitted values of $\beta$ and $\gamma$ for the Fe system are also physically reasonable, being slightly below one as is necessary for the clustering of point defects to be energetically favorable.
This is not the case for the CrCoNi and A-atom systems though;
both of these systems exhibit a $E_\mathrm{fit}$ that continually increases with dpa rather than converging around $0.5$ dpa, and $\gamma$ values that are well above the physically reasonable bound of $1.0$.
The second observation in particular suggests that the interstitial concentration is severely underestimated, with the model compensating for an unreasonably low value of $N_\mathrm{int}$ in Eq.\ \ref{eq:E_int} by elevating the value of $\gamma$.
Supposing from the Fe system that $\gamma$ should be $\about 0.9$, the interstitial concentration in the FCC systems appears to be underestimated by a factor of two to three at high dpa.
While it is true that the steady-state interstitial concentrations should not necessarily be comparable in BCC and FCC systems, Fig.\ \ref{fig:Point_defects} is also consistent with the interstitial concentrations in FCC systems being systematically underestimated by the same factor.
The source of the error is likely that the difference in atomic volumes of interstitials and atoms of the crystalline lattice is less pronounced in FCC than BCC systems, with interstitials in BCC Fe often adopting split-dumbbell configurations.

\begin{table}
    \caption{Parameters fit by the energy model that account for deviations of energy from isolated defects. $\beta$ is the coefficient for vacancies and $\gamma$ is for the interstitials.} 
    \begin{center}
        \begin{tabular}{@{}ccc@{}} \toprule
        & $\beta$ & $\gamma$ \\
        \midrule
        Fe & $0.879$ & $0.840$ \\
        CrCoNi & $0.454$ & $2.90$ \\
        A-atom & $0.970$ & $1.49$ \\
        \bottomrule
        \end{tabular} \label{table:1}
    \end{center} 
\end{table}


If the vacancy concentrations are relatively accurate and the errors in the interstitial concentrations are of a similar magnitude for the CrCoNi and A-atom systems, then the gap between the vacancy and interstitial concentrations would be significantly larger for the CrCoNi than for the A-atom system throughout the simulations, even at very low dpa.
If this is true, then that would have important implications for the effects of SRO and lattice distortions on the development of radiation damage.
Based on a conventional understanding of the point defect balance equations \cite{mansur1994theory,was2016fundamentals}, prior observations of CrCoNi's low-temperature radiation resistance \cite{velicsa2020temperature} imply that the gap between the vacancy and interstitial concentrations should be smaller for CrCoNi, encouraging point defect recombination and suppressing vacancy precipitation.
While this is superficially inconsistent with our results, observe that a higher sustained vacancy concentration does not necessarily lead to a higher susceptibility to voids.
Specifically, there is evidence that SRO and lattice distortions can increase the magnitude of energetic well depths for vacancies \cite{thomas2020vacancy}, effectively decreasing the vacancy-vacancy and vacancy-void capture radii and allowing higher vacancy concentrations to be sustained as compared to the A-atom system.
Given the magnitude of the apparent errors in the estimated vacancy and interstitial concentrations though, there is not strong evidence for this conclusion at present.

There are other qualifications relating to the CRA's limitations that should be made regarding the observed differences in point defect concentrations across the three material systems. 
One important factor absent in these simulations is the thermal spike associated with a collision cascade.
MPEAs have been found to have a shorter mean-free electron path and a lower thermal conductivity than traditional alloys, and it has been proposed that this could prolong the thermal spike and increase the extent of defect recombination \cite{zhang2015influence,zhang2021tunable}.
It is also worth considering the potential effects of nanotwinning, which has been experimentally observed to be an important cryogenic strengthening mechanism for CrCoNi \cite{ding2019real}. 
While the CRA did not induce any visible nanotwinning, this could be related to the relative sizes of the simulation cell and the critical nucleation event.
Nanotwins would not only affect the development of the dislocation network, but have shown the ability to capture or promote the transport of point defects to sinks in irradiated Cu \cite{chen2015damage}.

\section{Conclusion}
\label{sec:conclusion}

CRA simulations were conducted to investigate differences in the development of irradiated microstructures in BCC Fe, FCC CrCoNi, and FCC A-atom systems in the low temperature and high dose rate regime up to $2.0$ dpa.
The CrCoNi system developed the highest overall dislocation density and exhibited the ability to maintain that dislocation density even as the dislocation networks in the Fe and A-atom systems matured and simplified.
The higher stacking fault density in CrCoNi is likely related to the consistently higher density of partial dislocations relative to the other systems.
The successive insertion of Frenkel pairs entailed by the CRA mildly increased the degree of chemical short range order in CrCoNi relative to the initial random solid solution in a way that is consistent with other modeling and experimental studies.
A model was developed for the energy stored in the material defects, and strongly suggests that the interstitial concentrations for the FCC systems as estimated from the distribution of Voronoi cell volumes are lower than the actual effective concentrations by a factor of two to three.
It is also possible that the Wigner-Sietz method of identifying vacancies slightly overestimates the vacancy concentration for the FCC systems at high defect concentrations.
The same methods of point defect identification seem to be much more reliable for the BCC Fe system though, with the energy model closely following the simulation results.
Given the overwhelming importance of point defects to the development of radiation damage and the intense interest in CrCoNi and other FCC MPEAs for nuclear applications, our results reveal a critical need to either develop more robust ways to measure point defect concentrations in heavily-damaged FCC materials, or perhaps to reevaluate what is meant by a point defect in such materials.

\begin{acknowledgments}
JCS gratefully acknowledges partial support by the Nuclear Regulatory Commission (Award 31310019M0009) through the Advancing Scientific Careers to Enhance Nuclear Technologies (ASCENT) program at UC Davis.
This work was partially performed under the auspices of the U.S.\ Department of Energy by Lawrence Livermore National Laboratory under Contract DE-AC52-07NA27344.
\end{acknowledgments}

\section*{Author contributions}
\label{sec:contributions}

CS and JCS performed all the simulations and analysis. JKM conceptualized the project and developed the methodology. All authors collaborated to writing, reviewing, and editing.

\section*{Competing interests}
\label{sec:interests}

The authors declare no competing interests.

\appendix

\section{Convergence with system size}
\label{sec:appendix}

\begin{figure}
	\centering
    {\includegraphics[width=1.0\textwidth]{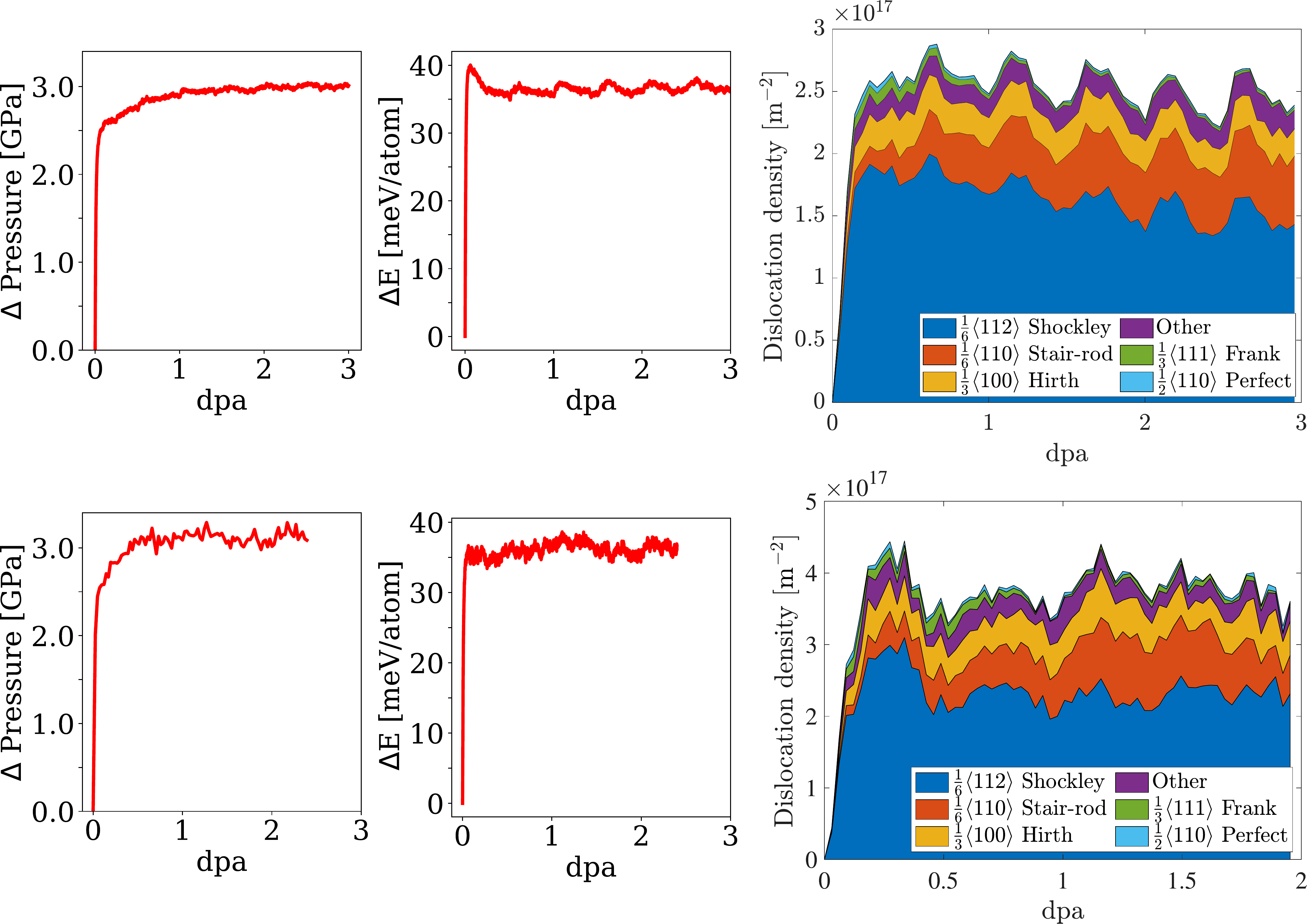}}
	\caption{Comparison of two simulations of equiatomic CrCoNi of different system sizes, with one million (one hundred thousand) atoms on the top (bottom) row.
 Included in this comparison are the change in pressure (left), change in potential energy (middle), and the dislocation density (right).}
	\label{fig:Convergence}
\end{figure}

As described in Sec.\ \ref{subsec:cra}, one simulation of equiatomic CrCoNi containing a total of $1\ 048\ 576$ atoms was conducted to $3.0\ \mathrm{dpa}$ to investigate the effect of system size. 
The behavior of this system is compared in Fig.\ \ref{fig:Convergence} (top row) with that of the smaller simulation containing a total of $131\ 072$ atoms (bottom row).
There is remarkable agreement between the two simulations, the main difference being that the plots for the larger simulation are smoother due to the larger number of atoms helping to average out the fluctuations.
Notice particularly that the pressure change and potential energy changes reach similar asymptotic values at similar dpa, that the dislocations in both are overwhelmingly Shockley partials, and that the overall dislocation density is relatively stable from $0.25$ dpa to the end of the simulations.

\bibliographystyle{apsrev4-2}
\bibliography{references}

\end{document}